\pdfoutput=1
\documentclass[11pt,fleqn]{article}
\usepackage{xcolor}
\usepackage[english]{babel}
\usepackage[utf8]{inputenc}
\usepackage[T1]{fontenc}
\usepackage{lmodern}
\usepackage{amssymb}
\usepackage{mathtools}
\usepackage[lmargin=1in,rmargin=1in,bottom=1in,top=1in,twoside=False]{geometry}
\usepackage{enumitem}
\usepackage{amsfonts}
\usepackage{amsmath}
\usepackage{amsthm}
\usepackage{complexity}
\usepackage{mathrsfs}
\usepackage{graphics}
\usepackage{amsfonts,amssymb} 
\usepackage{url}              
\usepackage[pagewise]{lineno}
\usepackage{microtype}
\usepackage[defaultlines=4,all]{nowidow}
\usepackage[many]{tcolorbox}
\tcbuselibrary{theorems}
\usepackage[colorinlistoftodos,prependcaption,textsize=scriptsize]{todonotes}
\usepackage{xspace}
\usepackage{makecell}
\usepackage{pifont}
\usepackage{bm}
\usepackage{scalerel}
\usepackage{tabularx}
\usepackage{afterpage}
\usepackage[hidelinks]{hyperref}
\usepackage[nameinlink]{cleveref}
\usepackage{thm-restate}
\usepackage{float}
\usepackage{xspace}
\usepackage{subcaption}
\usepackage{algorithm}
\usepackage{algpseudocode} 
\algtext*{EndIf}
\algtext*{EndFor}
\algtext*{EndWhile}

\definecolor{myPurple}{RGB}{230,230,255}
\definecolor{interesting}{RGB}{140,0,60}
\definecolor{brown}{RGB}{210,180,140}

\renewcommand{\epsilon}{\varepsilon}
\renewcommand{\phi}{\varphi}

\newcommand{\Gmc}{\ensuremath{\mathcal{G}}\xspace}

\theoremstyle{remark}

\Crefname{corollary}{Corollary}{Corollaries}
\Crefname{lemma}{Lemma}{Lemmas}
\Crefname{section}{Section}{Sections}

\newtheorem*{result*}{}
\newtheorem*{remark*}{Remark}

\begin{document}
\title{
Finding graph isomorphisms in heated spaces in almost no time\footnote{Source code for our implementation available at \url{https://github.com/s-najem/Isomorphism-Graph-Curvature}.}}
\author{Sara Najem \\ Department of Physics \\ Complexity and Network Science Cluster, CAMS \\ American University of Beirut, Lebanon \\ sn62@aub.edu.lb \and
Amer E. Mouawad \\ Department of Computer Science \\ Complexity and Network Science Cluster, CAMS \\ American University of Beirut, Lebanon \\ aa368@aub.edu.lb}
\date{}
\maketitle

\begin{abstract}
Graph isomorphism, the problem of determining whether two graphs encode the same combinatorial structure, has long challenged attempts at a purely structural resolution. We introduce a deterministic framework that approaches isomorphism through multi-scale diffusion coupled to geometry, establishing a connection between discrete spectral geometry and combinatorial algorithms. Each vertex is assigned a curvature-like signature derived from the short-time behavior of a (possibly fractional) graph Laplacian heat kernel, with dependence on spectral dimension. These signatures induce canonical vertex partitions that drive systematic vertex distinguishability and refinement.

Refinement proceeds in two stages. These diffusion-derived signatures provide an initial partition of the vertex set, which can then be systematically refined through additional structural probes. First, curvature-based signatures are aggregated to form equivalence classes of the original vertices. If non-singleton classes remain, refinement is strengthened through structured probing; selected vertices are temporarily augmented with controlled gadgets, and the induced partitions are compared to produce refined probe profiles. If termination has not been reached after this refinement stage, vertices are deterministically individualized through synchronized, permanent structural augmentation. These augmentations accumulate monotonically, yielding a geometry-guided individualization-refinement procedure.

The framework operates in deterministic polynomial time with respect to graph size and refinement parameters and constitutes a deterministic one-sided procedure; whenever it certifies isomorphism, the conclusion is correct. In \emph{all} experiments conducted to date, including random graphs, strongly regular graphs, and established benchmark instances, we did not observe any failures, although this remains an empirical observation rather than a provable guarantee.

Across these experiments, the framework consistently resolves instances that challenge classical spectral and refinement-based techniques. While not intended to compete directly with highly optimized solvers such as Nauty/Traces yet, the approach offers a complementary perspective based on geometric amplification rather than combinatorial search. Beyond isomorphism, the results suggest that diffusion-based curvature can serve as a structural descriptor of networks and provide a geometric perspective on vertex distinguishability in discrete systems.
\end{abstract}


\newpage
\tableofcontents
\newpage

\section{Introduction}

The graph isomorphism problem asks whether two finite graphs admit a bijection between their vertex sets that preserves adjacency. 
Despite decades of study, it occupies a singular position in complexity theory: it is neither known to admit a general polynomial-time algorithm nor believed to be NP-complete. 
Beyond its theoretical significance, graph isomorphism underlies applications ranging from chemical structure analysis to large-scale network comparison. The central challenge is to distinguish vertices in a canonical way without resorting to exhaustive search.

Most practical algorithms address this challenge through progressive refinement. Vertices are grouped into equivalence classes using combinatorial invariants, and these classes are iteratively refined using increasingly detailed structural information. If refinement separates all vertices into singletons, a canonical bijection is obtained and verified directly. This paradigm underlies influential procedures such as Weisfeiler--Leman refinements and their extensions, as well as highly optimized solvers such as Nauty/Traces~\cite{DBLP:journals/jsc/McKayP14}. However, purely combinatorial refinement can stabilize on highly regular graph families, leaving substantial ambiguity unresolved.

\begin{figure*}[ht]
  \centering
  \begin{subfigure}[c]{0.48\textwidth}
    \centering
    \includegraphics[width=1.1\linewidth]{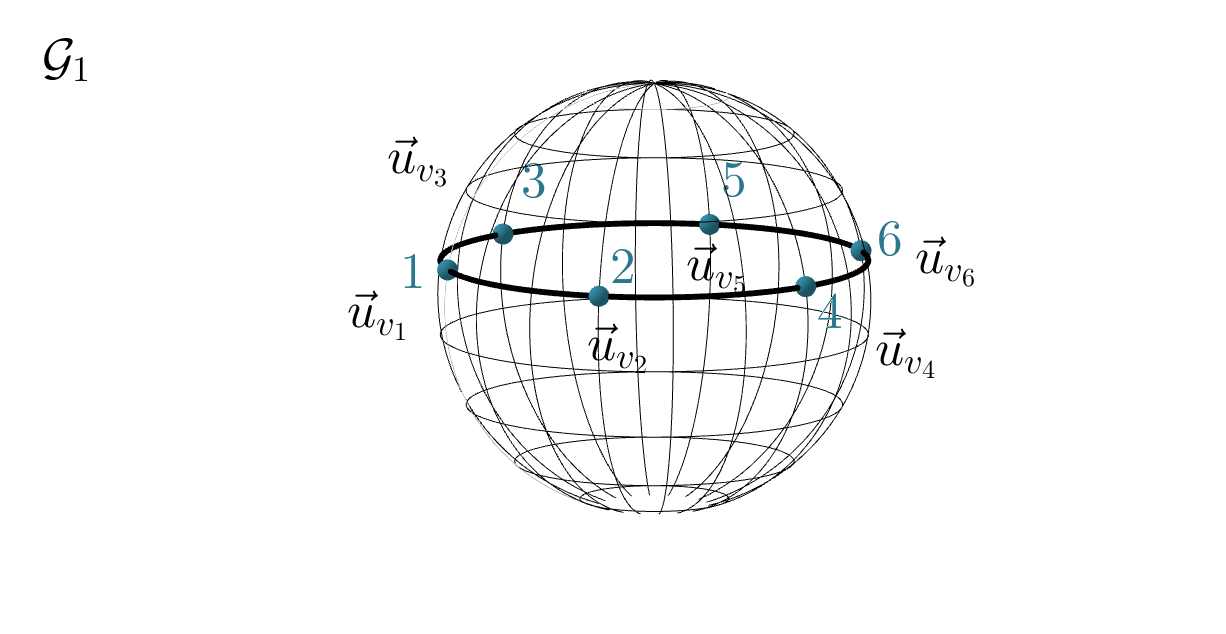}
    \caption{}
  \end{subfigure}
  \hfill
  \begin{subfigure}[c]{0.48\textwidth}
    \centering
    \includegraphics[width=1.1\linewidth]{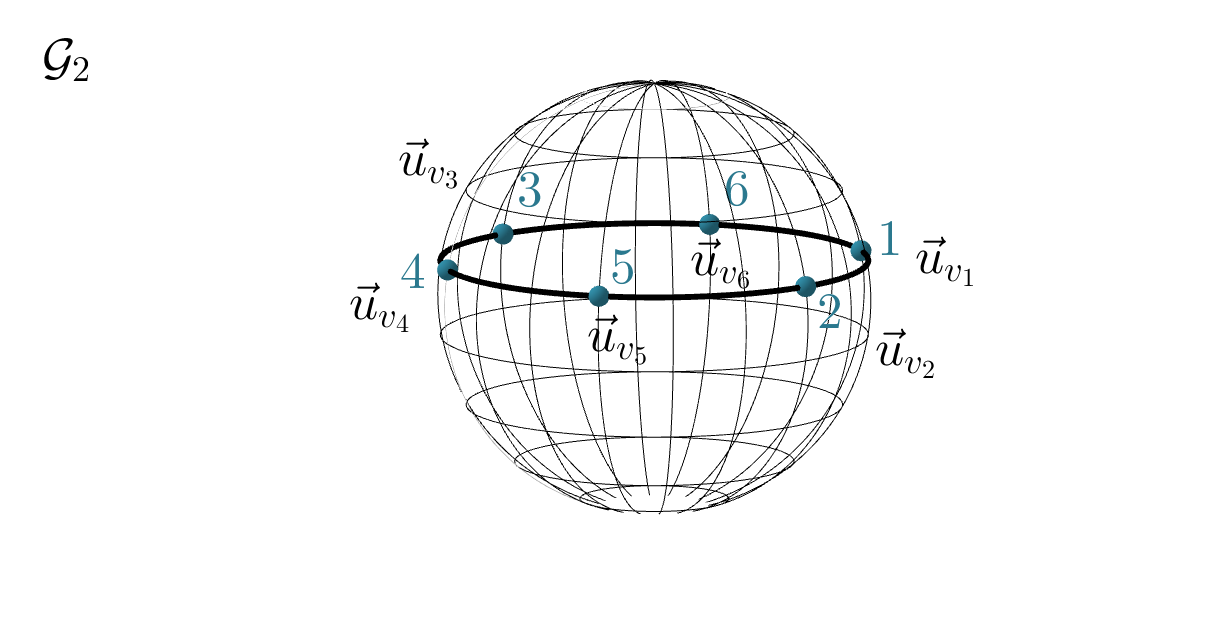}
    \caption{}
  \end{subfigure}
    \centering
  \begin{subfigure}[c]{0.48\textwidth}
    \centering
    \includegraphics[width=\linewidth]{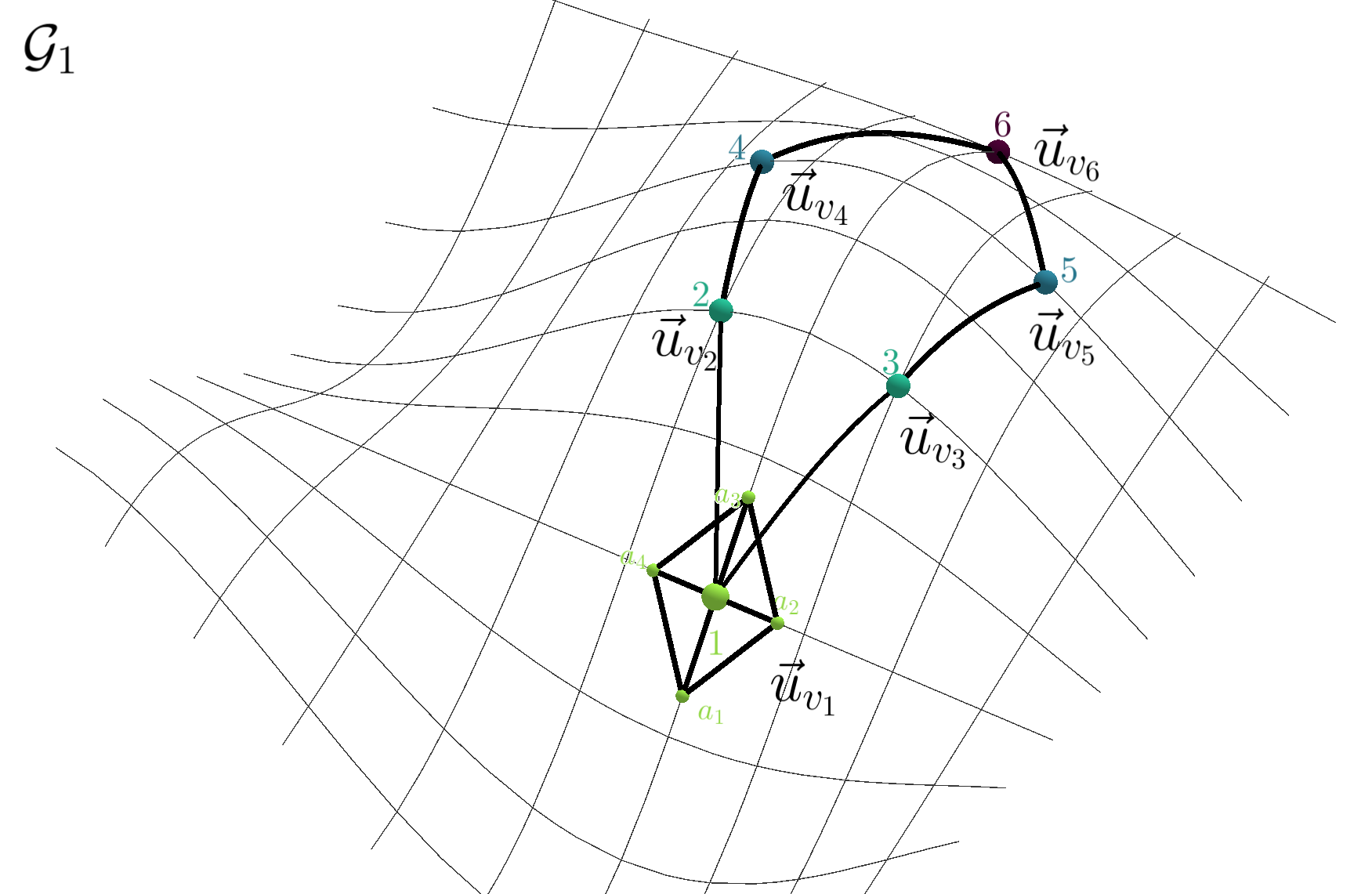}
    \caption{}
  \end{subfigure}
  \hfill
  \begin{subfigure}[c]{0.48\textwidth}
    \centering
    \includegraphics[width=1.1\linewidth]{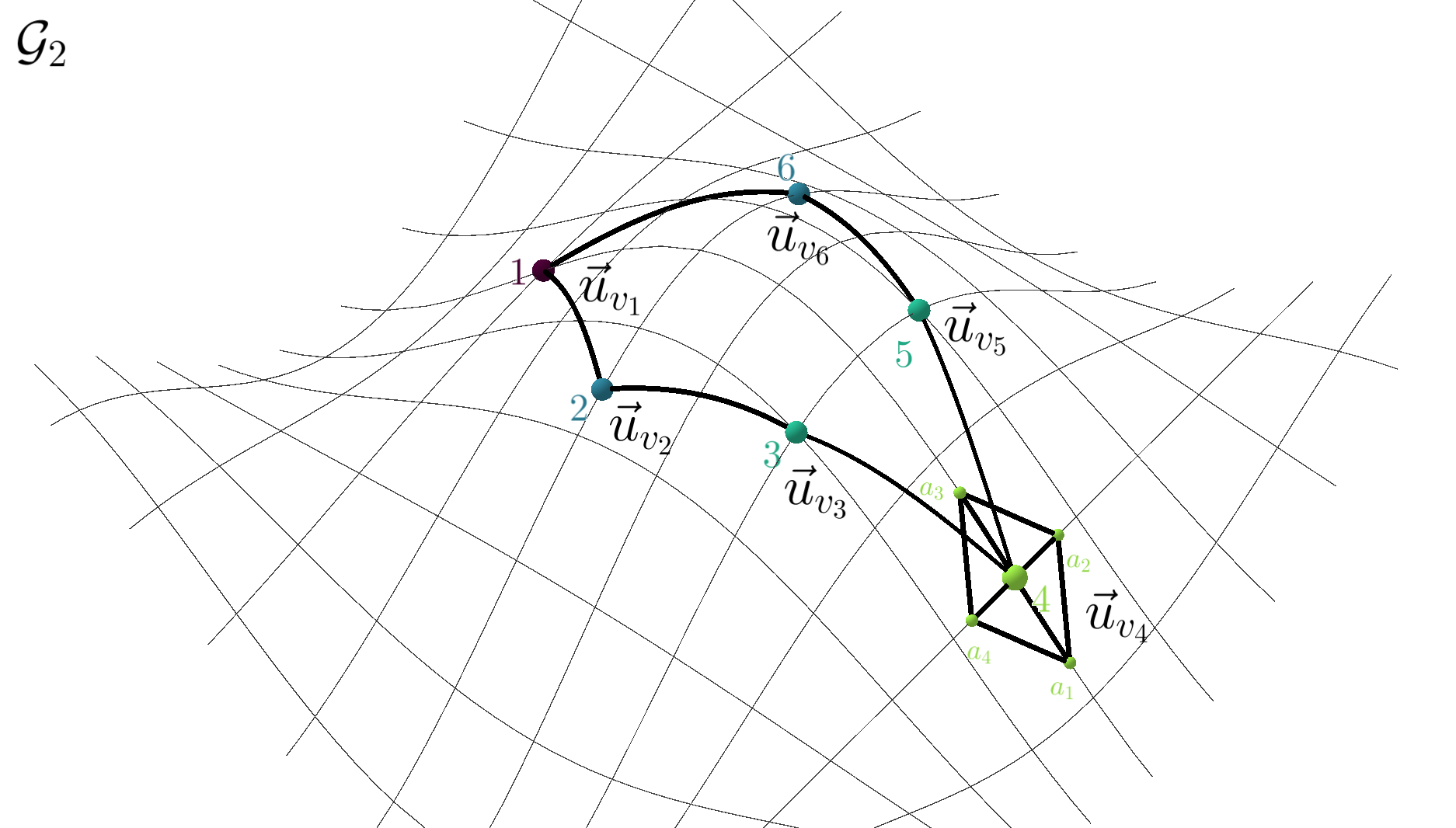}
    \caption{}
  \end{subfigure}
  \caption{The geometric signatures of the vertices $u_{v}$ of two cycles $\mathcal{G}_1$ and $\mathcal{G}_2$ are shown. The signatures are all equal, making the nodes indistinguishable and part of one equivalence class. Distinguishability is introduced by attaching a clique to vertex 1 in $\mathcal{G}_1$ and vertex 4 in $\mathcal{G}_2$, which allows their identification through their BFS-curvature signatures. The process is repeated iteratively on the remaining equivalence classes of indistinguishable vertices. }
  \label{fig:cyclesBRS}
\end{figure*}

Here we introduce a deterministic refinement framework driven by diffusion geometry. 
The short-time behavior of heat diffusion encodes multi-scale structural information that can enhance vertex distinguishability. 
Each vertex is assigned a curvature-like signature derived from a (possibly fractional) graph Laplacian heat kernel, scaled by its spectral dimension. 
These signatures induce canonical equivalence classes of the original vertices and provide a geometric description of  structure that propagates across the graph.

Refinement proceeds iteratively. First, diffusion-based signatures are aggregated over expanding neighborhoods to form vertex equivalence classes. 
If non-singleton classes remain, refinement is strengthened through structured probing; selected vertices from an  unresolved class are temporarily augmented 
with controlled gadgets, and the induced partitions of the probed graphs are compared. 
This probing phase refines the candidate class by grouping vertices according to their probe profiles, revealing distinctions not detected by diffusion alone.

Only after this refinement stage, if the partition still contains non-singleton classes that are not certified twin structures, does the procedure perform deterministic 
individualization. Guided by the refined probe partition, matched vertices in the two graphs are permanently augmented in a synchronized and monotone manner. 
Diffusion-based refinement then resumes on the enriched graphs. The procedure terminates when every original class is either a singleton or a twin class. 
In both cases, a vertex bijection consistent with the induced partition is constructed and explicitly verified on the original input graphs, 
guaranteeing one-sided correctness (see Figure~\ref{fig:cyclesBRS}).

We evaluate the framework on diverse graph families, including random graphs, strongly regular graphs, and established benchmark instances. 
Across all tested cases, including highly regular constructions that challenge classical spectral and refinement-only methods, the procedure successfully resolves isomorphism; we did not observe failure cases in our experiments, although this remains an empirical observation rather than a guarantee.

While the procedure is one-sided, in all experiments conducted to date (that manage to terminate using 128GB of RAM) we did not observe failure cases. 
A systematic comparison with state-of-the-art solvers such as Nauty/Traces, as well as a characterization of potential hard instances, remain important directions for future work.

More broadly, this work advances a geometric perspective on vertex distinguishability. 
It demonstrates that multi-scale diffusion geometry can serve as a systematic engine for refinement, 
providing a deterministic, geometry-guided complement to purely combinatorial or search-based approaches to graph isomorphism.

\section{Background and related work}

\subsection{Graph Laplacians, diffusion, and spectral geometry}

The graph Laplacian provides a fundamental analytic representation of a discrete structure. 
Given a simple undirected connected graph $\mathcal{G}=(V,E)$ with adjacency matrix $A$ and degree matrix $D$, the (combinatorial) Laplacian is defined as 
\[
L = D - A.
\]
The matrix $L$ is symmetric and positive semidefinite, with real, nonnegative eigenvalues. Its nullspace is spanned by indicator vectors of connected components, and the multiplicity of the zero eigenvalue equals the number of connected components \cite{Merris1994,Caughman2006}.

The spectrum of $L$ encodes structural information at multiple scales. The second-smallest eigenvalue (algebraic connectivity) reflects global expansion properties \cite{Fiedler1973,Alon1985}, while the full spectrum governs diffusion, random walks, and mixing processes \cite{Chung1997,Mohar1992}. These connections justify interpreting $L$ as a discrete analogue of the Laplace--Beltrami operator on a Riemannian manifold.

Diffusion on a graph is governed by the heat equation
\[
\frac{\partial u}{\partial t} = -Lu,
\]
whose solution is given by the heat kernel
$
K(t)=e^{-tL}.
$
The heat kernel interpolates between local and global structure; small values of $t$ probe immediate neighborhoods, whereas larger values reflect long-range connectivity \cite{MasudaPorterLambiotte2020}. In continuous geometry, short-time heat asymptotics encode intrinsic quantities such as dimension and curvature \cite{Petersen2016}. This motivates extracting analogous geometric descriptors from discrete diffusion processes.

Fractional Laplacians extend this framework by spectrally scaling eigenvalues $\lambda_i \mapsto \lambda_i^s$ for $s>0$, to account for non-locality. Such scaling modifies diffusion rates across frequency bands and has been studied in anomalous transport and complex networks \cite{AlexanderOrbach1982,BurioniCassi1996,Rozenfeld2010deterministic}. Closely related is the notion of spectral dimension, which characterizes the asymptotic scaling of eigenvalue density near zero. In the present work the spectral dimension plays an important role in defining heat-kernel scaling across graph families.

\subsection{Spectral methods and graph isomorphism}

Because the Laplacian is invariant under vertex relabeling, spectral quantities are natural graph invariants. If two graphs are isomorphic, their Laplacians are related by conjugation with a permutation matrix and therefore share identical spectra \cite{Merris1994}. Spectral mismatch thus certifies non-isomorphism.

However, spectral equality does not imply isomorphism. Infinite families of non-isomorphic cospectral graphs are known \cite{Merris1994}, demonstrating that eigenvalues alone are insufficient to capture full combinatorial structure. This limitation parallels classical questions in spectral geometry concerning the extent to which geometric structure is determined by eigenvalues.

Despite these limitations, spectral information is widely used in practical isomorphism testing. Eigenvectors provide geometric embeddings of vertices, and heat-kernel-based descriptors yield multi-scale vertex signatures \cite{Reuter2006,Sun2009,Raviv2013}. More recent approaches aggregate spectral information across eigenmodes to increase discriminative power \cite{DjimaYim2025}. These developments indicate that while raw spectra are incomplete invariants, structured spectral descriptors can serve as powerful refinement tools.

\subsection{Combinatorial refinement and algorithmic complexity}

Combinatorial refinement is central to modern isomorphism solvers. The Weisfeiler--Leman (WL) hierarchy iteratively refines vertex colors based on neighborhood structure and is highly effective in practice~\cite{Weisfeiler1968}. Practical systems such as Nauty and Traces combine refinement with search and sophisticated partitioning strategies \cite{DBLP:journals/jsc/McKayP14}. Nevertheless, WL refinement can stabilize on highly regular graph families, including strongly regular graphs~\cite{babai1980canonical,uehara2005tractable}, leaving substantial ambiguity unresolved.

From a complexity perspective, graph isomorphism is polynomial-time solvable for numerous restricted classes, including trees \cite{aho1974design}, bounded-degree graphs \cite{luks1982isomorphism}, planar graphs \cite{hopcroft1974linear}, and bounded treewidth graphs \cite{grohe2015graph}. Babai’s quasipolynomial-time algorithm established a general upper bound of $n^{O(\log^c n)}$ for all graphs \cite{babai2016graph}. However, the existence of a deterministic polynomial-time algorithm for general graph isomorphism remains an open question.

\subsection{Diffusion geometry as a refinement engine}

The preceding discussion highlights a recurring obstacle: persistent indistinguishability. Both combinatorial refinement and coarse spectral invariants may fail when  structure is highly regular. The difficulty lies not in graph size but in the inability of refinement mechanisms to propagate distinguishing information. 

This observation motivates integrating diffusion and geometry directly into the refinement process. Short-time heat diffusion encodes subtle structural distinctions that may be invisible to purely combinatorial statistics. By organizing diffusion-derived signatures into deterministic refinement steps, and strengthening refinement via controlled probing when necessary, one can attempt to transform geometric information into a systematic refinement mechanism for vertex distinguishability.

The present work develops this idea into a concrete algorithmic framework. Diffusion-based signatures define initial equivalence classes; structured probing refines unresolved classes; and deterministic (geometric) individualization applies synchronized augmentations when refinement alone is insufficient. In contrast to approaches that use spectral data only as auxiliary invariants, diffusion and geometry here serve as a primary driver of refinement.

\section{The base algorithm for computing curvatures}

We begin by describing the geometric core of our framework. 
At this stage, the algorithm operates purely at the level of  diffusion and its relation to geometry. 
It does not construct vertex correspondences or refine equivalence classes. 
Instead, it extracts canonical curvature-like descriptors from the graph Laplacian that serve as the foundational invariants for all subsequent refinement and individualization stages.

\subsection{Laplacian and heat kernel}

Let $\mathcal{G}=(V,E)$ be a simple, undirected, connected graph with $|V|=n$. 
Let $A$ denote its adjacency matrix and $D$ its degree matrix. 
The combinatorial Laplacian is defined by
$
L = D - A.
$
The matrix $L$ is symmetric and positive semidefinite and admits an orthonormal eigenbasis \cite{Merris1994,Mohar1992,Chung1997}. 
Let
$
L = V^\top \Lambda V,
$
where $\Lambda=\mathrm{diag}(\lambda_1,\dots,\lambda_n)$ and $V$ contains the corresponding eigenvectors.

Diffusion on the graph is governed by the heat equation
\[
\frac{\partial u}{\partial t} = -Lu,
\]
whose solution is given by the heat kernel
$
K(t)=e^{-tL}=V^\top e^{-t\Lambda} V.
$
The diagonal entries $K(i,i,t)$ represent heat-return probabilities at vertex $i$.
These quantities encode diffusion behavior across scales and form the basis of our geometric descriptors.
Diffusion processes on networks and their spectral interpretation are well studied in both graph theory and network science \cite{Chung1997,MasudaRochaEtAl2017,MasudaPorterLambiotte2020}.

Because $L$ is defined canonically from $\mathcal{G}$, the heat kernel is equivariant under vertex relabeling. 
Any vertex permutation conjugates $L$ and therefore transforms $K(t)$ accordingly, ensuring that all derived quantities are graph invariants.

\subsection{Spectral dimension and short-time regime}

The short-time behavior of the heat kernel depends on how the Laplacian spectrum scales near zero.
To differentiate diffusion across graph families, we estimate the spectral dimension $d_s$, defined via the asymptotic scaling of the cumulative eigenvalue density \cite{torres2020simplicial}, or equivalently the counting function
\[
\rho(\lambda' < \lambda) \sim \lambda^{d_s/2}
\quad \text{as } \lambda \to 0.
\]
This notion originated in the study of diffusion on fractals and disordered media \cite{AlexanderOrbach1982,BurioniCassi1996} and has been extended to hierarchical and complex networks \cite{Rozenfeld2007flowers,Rozenfeld2010deterministic,Hwang2010,Bianconi2019}.

In practice, $d_s$ is estimated by linear regression on logarithmically scaled portions of the spectrum. 
The estimate is not itself used as a vertex-level invariant. 
Rather, it determines an appropriate short-time window in which diffusion primarily reflects local geometry rather than global mixing.

Time samples are chosen logarithmically within this short-time regime. In order to identify this short time interval, the trace of the heat kernel, defined as $\text{Tr}(K)$, 
is evaluated on the interval $[1/\lambda_{max},1/\lambda_{min}]$. The upper limit of the short time interval $t^+$ is defined such that it maximizes the adjusted $\text{-}R^2$ of the fit of $\log{\text{Tr}(K)}$ versus time, while its lower limit is given by $1/\lambda_{max}$. 
This ensures stable coverage of multiple local scales while avoiding numerical instability at extremely small $t$ and domination by global diffusion at larger $t$.

\subsection{Curvature extraction from short-time expansion}
On a manifold, in continuous Riemannian geometry, and close to  $t^+ \rightarrow 0$,  the heat kernel expansion is given by:
\begin{equation}\label{heatkern}
K(x, y, t) \sim \frac{1}{(4\pi t)^{n/2}} \exp\left( -\frac{d(x,y)^2}{4t} \right) \sum_{k=1}^\infty u_k(x, y) t^k,
\end{equation}
with $n$ denoting the dimension of the manifold, $d(x,y)$ being the distance between arbitrary points in space and $t$ is time. 
The diagonal heat kernel admits a short-time asymptotic expansion
\[
K(x,x,t) \sim \frac{1}{(4\pi t)^{n/2}} 
\sum_{k=0}^{\infty} u_k(x)\, t^k,
\]
where the Minakshisundaram-Pleijel coefficients $u_k(x)$ encode geometric information such as 
curvature~\cite{Petersen2016,minakshisundaram1949some,chavel1984eigenvalues,najem2025geometric}. It is well established that scalar curvature $R(x)$, 
which is the curvature at a given point in space, is given as the second term in the expansion, i.e., the coefficient that corresponds to $k=1$.

Motivated by this analogy and by diffusion-based signatures in geometry processing \cite{Sun2009,Reuter2006,Raviv2013}, and taking $x\rightarrow i$, where the indices $i$ denote the vertices' identities, we model the short-time behavior of $K(i,i,t)$ by fitting a truncated expansion
\[
K(i,i,t) \approx 
t^{-d_s/2} \sum_{k=0}^{N} u_k(i)\, t^k
\]
over the selected short-time interval.

The coefficients $u_k(i)$ are obtained via least-squares regression and serve as curvature-like descriptors of vertex $i$.
Lower-order terms capture coarse diffusion geometry, while higher-order coefficients encode increasingly fine structural detail.
Because they are derived canonically from the Laplacian spectrum and eigenvectors, these descriptors are equivariant under vertex relabeling.

For algorithmic robustness, coefficients are discretized via fixed scaling and rounding, yielding deterministic signatures suitable for consistent cross-graph comparison. 

\subsection{Resolving indistinguishable configurations and fractional scaling}

Purely  combinatorial refinement can fail in highly regular or indistinguishable configurations. 
A simple example is a long cycle ($n \geq 200$) with a small clique, i.e. a triangle, attached on some arbitrary vertex. 
Along the cycle, all vertices share identical  neighborhood structure.
Vertices far away from the clique may remain indistinguishable under low-order combinatorial statistics.
Classical Laplacian eigenvalues as well as curvature values, while informative, may also fail to provide sufficient vertex-level separation in such settings \cite{Merris1994,Raviv2013}.

This illustrates a broader limitation: under classical diffusion $e^{-tL}$, contributions from different spectral scales are fixed.
Low-frequency eigenmodes dominate long-time behavior, while high-frequency modes decay rapidly.
In graphs with extended regular regions, this can produce diffusion profiles that remain too uniform to separate vertices early in refinement.

To address this limitation, we also consider a spectrally scaled diffusion operator corresponding to a fractional graph Laplacian.
Fractional Laplacians have been studied both theoretically \cite{Zhang2025FLO,Zhang2025FV} and in applications to nonlocal network dynamics and graph learning \cite{Benzi2020,Maskey2023,Weihs2024}.
Given eigenvalues $\{\lambda_i\}$, we define modified eigenvalues
$
\mu_i = \lambda_i^{\,s},
$
for an exponent $s>0$, and define the scaled heat kernel
\[
K_s(t) = V^\top e^{-t\,\mathrm{diag}(\mu_1,\dots,\mu_n)} V.
\]

When $s=1$, classical diffusion is recovered.
For $s\neq 1$, spectral bands are reweighted.
If $s<1$, low-frequency modes are relatively amplified, enhancing long-range sensitivity.
If $s>1$, higher-frequency components are emphasized, increasing sensitivity to fine irregularities.

The exponent $s$ is chosen in relation to the estimated spectral dimension, normalizing diffusion behavior across graph families with different effective dimensionality.
Importantly, fractional scaling remains canonical, i.e., it modifies spectral weighting but introduces no external labeling or bias. 

Returning to the cycle-with-clique example, fractional scaling alters heat-return behavior.
Under suitable scaling, deviations become detectable at lower polynomial orders in the short-time expansion than in the classical case.
Thus vertices that are initially indistinguishable can separate purely through diffusion geometry.

\subsection{Role of the base geometric pipeline}

The base pipeline associates to each vertex $i$ a finite vector
\[
\mathbf{u}(i) = (u_0(i),\dots,u_K(i))
\]
derived from short-time diffusion under (possibly fractional) spectral scaling.

These vectors provide canonical, multi-scale geometric fingerprints of vertices.
While insufficient by themselves to resolve graph isomorphism in highly regular graphs, they establish the geometric quantities that drive all subsequent refinement, probing, and individualization stages.
In the next sections, we show how these descriptors are aggregated and systematically amplified to form a deterministic refinement hierarchy.

\section{From local curvature to BFS-curvature signatures}

The curvature coefficients extracted by the base algorithm provide ``local'' geometric descriptors for individual vertices. 
However, local diffusion geometry alone is often insufficient to distinguish vertices in graphs with repetitive or highly regular structure. 
To propagate local information into a structured neighborhood context, we aggregate curvature data across graph neighborhoods using a breadth-first search (BFS) hierarchy.

This aggregation produces BFS-curvature signatures, which encode multi-scale diffusion geometry around each vertex in a canonical, distance-stratified manner. 
These signatures constitute the first refinement mechanism of the algorithm, transforming vertex-level descriptors into structured neighborhood-level invariants.

\subsection{Definition of BFS-curvature signatures}

Let $\mathcal{G} = (V,E)$ denote the current working graph. 
The working graph may contain auxiliary vertices introduced during preprocessing or controlled perturbations. 
However, signatures are computed and compared only for the original vertices of the input graph(s).

Let $u_k(i)$ denote the $k$-th heat-kernel coefficient at vertex $i$, and let
\[
\mathbf{u}_K(i) = (u_0(i), u_1(i), \dots, u_K(i))
\]
be the truncated curvature vector at order $K$. 
For deterministic comparison, each coefficient vector is discretized using a fixed scaling and rounding scheme.

For a vertex $v$, perform a BFS rooted at $v$ in the working graph. 
This partitions vertices into layers
\[
\mathcal{L}_0(v) = \{v\}, \quad
\mathcal{L}_1(v), \quad
\mathcal{L}_2(v), \quad \dots,
\]
where $\mathcal{L}_r(v)$ consists of vertices at graph distance $r$ from $v$.

The BFS-curvature signature of $v$ at truncation order $K$ is defined as
\[
\mathrm{BCS}_K(v)
=
\Bigl(
\mathbf{u}_K(v),
\;\mathrm{sort}\{\mathbf{u}_K(w) : w \in \mathcal{L}_1(v)\},
\;\mathrm{sort}\{\mathbf{u}_K(w) : w \in \mathcal{L}_2(v)\},
\;\dots
\Bigr),
\]
where sorting within each layer is performed lexicographically.

The layer index encodes discrete distance from the root, while lexicographic sorting removes dependence on vertex labeling within each layer. 
Consequently, $\mathrm{BCS}_K(v)$ is canonically defined from graph structure and diffusion data and is equivariant under vertex relabeling.

\subsection{Multi-scale geometric interpretation}

The BFS-curvature signature can be interpreted as a discrete multi-scale geometric profile. 
The root contributes its intrinsic diffusion geometry, and successive BFS layers describe how this geometry interacts with increasingly distant neighborhoods.

Two vertices may share identical leading-order curvature coefficients yet differ in the organization of their surrounding diffusion profiles. 
By stratifying curvature data according to graph distance, BFS-curvature signatures capture both coefficient depth and spatial arrangement. 
As the truncation order $K$ increases, the signatures incorporate increasingly fine  diffusion information. 
As BFS depth increases, the spatial extent of aggregation grows outward from the root. 
Together, these parameters provide two independent axes of geometric resolution.

The BFS-curvature signature is thus viewed as a discrete analogue of the  curvature profile of 
a point on a Riemannian manifold. In differential geometry, the full collection of curvature invariants (and their derivatives) at 
a point determines the  metric up to isometry (Cartan–Ambrose–Hicks theorem). 
Similarly, $\mathrm{BCS}(v)$ encodes the geometry of the metric-ball around $v$ in the graph. 
Matching BFS-curvature signatures across two graphs is therefore analogous to constructing an isometry: if every vertex in $\Gmc_1$ has an 
identical signature to some vertex in $\Gmc_2$, then the two graphs share the same discrete metric structure. 
In effect, sorting and aligning the signatures from $\Gmc_1$ and $\Gmc_2$ ``enforces'' a discrete isometry between the graphs.

\subsection{Induced equivalence classes}

For a fixed truncation order $K$, BFS-curvature signatures induce an equivalence relation on the original vertex set:
\[
v \sim_K w \quad \text{if and only if} \quad \mathrm{BCS}_K(v) = \mathrm{BCS}_K(w).
\]
This defines a partition of the original vertices into signature classes.

Given two graphs $\mathcal{G}_1$ and $\mathcal{G}_2$, the multisets of signatures at order $K$ are compared. 
If a signature appears with different multiplicities in the two graphs, they are deemed not isomorphic. 
If the induced partitions contain only singleton classes, a canonical bijection between vertices is determined by matching identical signatures.

If non-singleton classes persist and do not correspond to certified twin structures, the available geometric resolution is insufficient to fully distinguish vertices. 
In that case, refinement must continue.

\subsection{Increasing geometric resolution}

Geometric resolution is increased deterministically by enlarging one or both of the following parameters:

\begin{itemize}
\item the truncation order $K$ of the short-time expansion, and
\item the BFS depth used in constructing the signature.
\end{itemize}

Both parameters are increased in a controlled, bounded manner. 
At each stage, the induced partition is recomputed and compared across graphs. 
Only when geometric refinement stabilizes without achieving termination does the algorithm proceed to structured probing. In our implementation, 
we let $K$ and BFS depth be parameters, where $K$ is set to two plus the maximum degree in the original graphs by default and BFS depth is set to infinity, i.e., BFS includes all vertices of the graph (since we assume connected graphs). 
We note here that disconnected graphs can easily be handled by adding a universal vertex, i.e., a vertex connected to all vertices of the input graphs. 

\subsection{Role in the overall algorithm}

BFS-curvature signatures form the primary intrinsic refinement mechanism of the algorithm. 
They propagate vertex-level diffusion geometry into structured, multi-scale invariants that refine vertex equivalence classes without modifying the graph.

Temporary probing, described in the next section, strengthens this refinement when intrinsic diffusion geometry alone is insufficient. 
Permanent individualization is invoked only after refinement, including probing, fails to achieve termination. 
This separation ensures that geometric information is fully exploited before structural augmentation is introduced.

\section{Geometric normalization, refinement, and vertex individualization}

BFS-curvature signatures provide a structured, multi-scale description of diffusion geometry around each vertex. 
In many graphs, these signatures refine all equivalence classes to singletons. 
However, at any fixed geometric resolution, multiple vertices may remain indistinguishable. 
Such degeneracies do not necessarily indicate the presence of automorphisms, nor do they imply failure of the method. 
They indicate only that the currently available geometric information is insufficient to separate certain vertices.

To address this, the algorithm applies a deterministic refinement hierarchy. 
Refinement is performed symmetrically on both input graphs and is interpreted exclusively through induced equivalence classes on the original vertex sets.

\subsection{Geometric normalization via subdivision}

Stable extraction of short-time diffusion coefficients requires an appropriate geometric scale. 
In certain graph families, particularly those with small diameter or strong regularity, diffusion may mix too rapidly to admit stable regression of curvature coefficients.

To regulate this behavior, the algorithm may apply bounded edge subdivision as a geometric normalization step. 
Each edge is replaced by a short path of fixed length, producing a subdivided graph. 
Subdivision is applied symmetrically to both input graphs and preserves isomorphism.

Subdivision vertices participate in diffusion and BFS-signature computations but are treated as auxiliary. 
They are excluded from equivalence-class comparison and from the final vertex correspondence. 
The subdivision depth is deterministically bounded by a polynomial in $n$.

\subsection{Structured probing as refinement}

If increasing coefficient depth and BFS aggregation fail to separate a non-singleton equivalence class $\mathcal{C}$ of original vertices, refinement is strengthened 
through structured probing.

Probing is temporary and diagnostic, and does not affect the final correctness of the procedure. 
For a selected vertex $u \in \mathcal{C}$, augmented graphs are constructed in which carefully designed gadgets are attached to $u$ together with other vertices in $\mathcal{G}$. 
The augmented graphs are processed through the same curvature and BFS-signature pipeline, and only the induced partitions on original vertices are retained. 
Auxiliary gadget vertices are discarded after each probe.

The primary probing mechanism operates on ordered vertex pairs. 
For each pair $(u,v)$ with $u \in \mathcal{C}$, $v \in \mathcal{G}$, and $u \neq v$, symmetric attachments are introduced in both graphs. 
These attachments combine a shared clique structure with private distinguishing components whose parameters are varied deterministically. 
Both orientations are evaluated to preserve symmetry under labeling.

For a fixed $u$, the multiset of induced refinement outcomes against all $v \neq u$ defines a probe profile. 
Vertices in $\mathcal{C}$ are partitioned according to equality of probe profiles. 
If this partition strictly refines $\mathcal{C}$ while remaining consistent across the two graphs, the refinement is accepted.

If pair probing does not refine the class, the algorithm escalates to triplet probing. 
Triples $(u,v,w)$ are augmented using three distinct private attachments, again evaluated under symmetric orientations. 
Triplet probing incorporates ternary structural relationships and increases discriminative power.

Probing refines equivalence classes but does not permanently modify the graphs. 
Permanent augmentation occurs only after refinement has produced a consistent partition that does not yet satisfy termination conditions.

\subsection{Deterministic individualization}

After refinement (including probing) is applied, the induced partition on original vertices is evaluated. 
If all classes are singletons or correspond to certified twin structures, the algorithm terminates. 
Otherwise, the algorithm performs deterministic vertex individualization.

Individualization consists of synchronized, permanent structural augmentation of matched vertices in the two graphs. 
Selected vertices, guided by the refined partition, are augmented with clique attachments or related structures. 
These augmentations accumulate monotonically across rounds and alter the working graphs for subsequent refinement.

Following individualization, geometric refinement (coefficient depth, BFS signatures, and probing) resumes on the enriched graphs.

\subsection{Refinement strategy and termination}

Refinement proceeds along three deterministically bounded axes:
\begin{enumerate}
\item heat-kernel coefficient depth,
\item BFS aggregation radius,
\item probing complexity (pair followed by triplet probing).
\end{enumerate}

At each stage, equivalence classes on original vertices are recomputed and compared across the two graphs. 
Refinement is accepted only if the induced partitions remain consistent across the two graphs. 
The algorithm terminates when:
\begin{itemize}
\item all original equivalence classes are singletons, or
\item the remaining non-singleton classes correspond to provably symmetric twin vertices (true or false twins).
\end{itemize}

In either case, a vertex bijection consistent with the induced partition is constructed and explicitly verified on the original, unmodified graphs.

All refinement parameters, including coefficient depth, BFS aggregation, subdivision depth, and probing complexity, are bounded by deterministic polynomials in $n$, ensuring that the overall procedure runs in polynomial time.

\subsection{Interpretation}

Refinement and probing are used solely to extract distinguishing information. 
Any candidate bijection induced by singleton or twin classes is explicitly verified on the original graphs by checking adjacency preservation. 
Consequently, the procedure is one-sidedly correct; it never declares two non-isomorphic graphs to be isomorphic, and any positive identification is explicitly verified on the original inputs. 
Failure to produce a verified bijection indicates only that the instance could not be resolved within the prescribed refinement bounds and does not imply isomorphism.

The refinement hierarchy can be viewed as progressively sharpening geometric resolution. 
At low resolution, diffusion may render distinct vertices indistinguishable. 
By increasing spectral depth, expanding spatial aggregation, and incorporating higher-order probing when necessary, the algorithm systematically amplifies latent asymmetries while maintaining determinism and polynomial bounds.

\section{The complete algorithm for graph isomorphism}

We now describe the complete deterministic procedure for graph isomorphism.
The algorithm integrates diffusion-based curvature extraction, BFS-curvature aggregation, structured probing, and deterministic vertex individualization into a single refinement hierarchy.
At every stage, both input graphs are processed symmetrically.
All decisions are based solely on equivalence-class comparisons of original vertices, and any candidate correspondence is accepted only after explicit verification on the original graphs.

\subsection{Algorithm overview}

Let $\mathcal{G}_1=(V_1,E_1)$ and $\mathcal{G}_2=(V_2,E_2)$ be connected graphs with $|V_1|=|V_2|=n$.
We identify the original vertices with $\{0,1,\dots,n-1\}$.
The algorithm maintains working graphs $\widetilde{\mathcal{G}}_1$ and $\widetilde{\mathcal{G}}_2$, which may contain auxiliary vertices introduced by a one-time symmetric normalization (edge subdivision) step and by permanent individualization gadgets.
Structured probing is performed on temporary copies and does not permanently modify $\widetilde{\mathcal{G}}_1,\widetilde{\mathcal{G}}_2$.

The procedure consists of a deterministic refinement loop.
Algorithm~\ref{alg:gi-core} gives the high-level pseudocode; in what follows we define each component precisely.

\begin{algorithm}[t]
\caption{Proposed algorithm for graph isomorphism}
\label{alg:gi-core}
\begin{algorithmic}[1]
\Require Connected graphs $\mathcal{G}_1,\mathcal{G}_2$ on $n$ vertices
\Ensure \textbf{isomorphic} with verified bijection, or \textbf{fail}


\State Determine minimal subdivision level $s$ satisfying bounds and non-degenerate spectral dimension
\State $\widetilde{\mathcal{G}}_1 \gets \mathrm{SubdivideEveryEdge}(\widetilde{\mathcal{G}}_1,s)$
\State $\widetilde{\mathcal{G}}_2 \gets \mathrm{SubdivideEveryEdge}(\widetilde{\mathcal{G}}_2,s)$

\State $q \gets q_0$ \Comment{initial gadget size}

\While{true}

  \State $\Pi_1 \gets \mathrm{SignaturePartition}(\widetilde{\mathcal{G}}_1)$ \Comment{original-only signatures}
  \State $\Pi_2 \gets \mathrm{SignaturePartition}(\widetilde{\mathcal{G}}_2)$

  \If{$\Pi_1,\Pi_2$ disagree in signature multiplicities}
    \State \Return \textbf{fail}
  \EndIf

  \If{$\Pi_1,\Pi_2$ are all singletons, or all non-singletons are matched twin classes}
    \State $f \gets \mathrm{CanonicalBijectionFromClasses}(\Pi_1,\Pi_2)$
    \If{$\mathrm{VerifyAdjacency}(\mathcal{G}_1,\mathcal{G}_2,f)$}
      \State \Return \textbf{isomorphic}
    \Else
      \State \Return \textbf{fail}
    \EndIf
  \EndIf

  \State $(\mathcal{C}_1,\mathcal{C}_2) \gets \mathrm{MatchedNonSingletonClass}(\Pi_1,\Pi_2)$

  \State $R_1 \gets \mathrm{ProbeRefine}(\widetilde{\mathcal{G}}_1,\mathcal{C}_1,q)$ \Comment{pair then triplet}
  \State $R_2 \gets \mathrm{ProbeRefine}(\widetilde{\mathcal{G}}_2,\mathcal{C}_2,q)$

  \State $M \gets \mathrm{MatchRefinedGroupsByProfiles}(R_1,R_2)$

  \If{no singleton groups in both $R_1$ and $R_2$}
    \State $(A,B) \gets \mathrm{SmallestMatchedGroupPair}(M)$
    \State $(u,v) \gets \mathrm{ChooseRepresentatives}(A,B)$
    \State $\widetilde{\mathcal{G}}_1 \gets \mathrm{AttachClique}(\widetilde{\mathcal{G}}_1,u,q)$
    \State $\widetilde{\mathcal{G}}_2 \gets \mathrm{AttachClique}(\widetilde{\mathcal{G}}_2,v,q)$
    \State $q \gets q+1$
  \Else
    \For{each matched singleton pair $(u,v)$ in deterministic order}
      \State $\widetilde{\mathcal{G}}_1 \gets \mathrm{AttachClique}(\widetilde{\mathcal{G}}_1,u,q)$
      \State $\widetilde{\mathcal{G}}_2 \gets \mathrm{AttachClique}(\widetilde{\mathcal{G}}_2,v,q)$
      \State $q \gets q+1$
    \EndFor
  \EndIf

\EndWhile
\end{algorithmic}
\end{algorithm}

\paragraph{Step 1: Initialization and symmetric normalization.}
The algorithm first checks connectivity and fixed input bounds.
If required, it applies a bounded symmetric normalization by subdividing every edge a common number of times $s\ge 0$.
Starting from a prescribed $s_0$, it increases $s$ until both subdivided graphs (i) lie within an allowed vertex-count window and (ii) yield a non-degenerate spectral-dimension estimate.
This subdivision level is fixed and the resulting graphs become the initial working graphs $\widetilde{\mathcal{G}}_1,\widetilde{\mathcal{G}}_2$.

\paragraph{Step 2: Curvature computation and signature partitioning.}
On each refinement round and for each $i\in\{1,2\}$:
\begin{itemize}
\item compute the Laplacian spectrum of $\widetilde{\mathcal{G}}_i$;
\item estimate the spectral dimension;
\item compute a fixed number of short-time curvature coefficients per vertex;
\item construct BFS-curvature signatures for original vertices only (as defined previously).
\end{itemize}
This induces a partition $\Pi_i$ of $\{0,\dots,n-1\}$ into signature classes.

The partitions are compared across graphs by signature type and multiplicity.
If some signature occurs with different multiplicity in $\Pi_1$ and $\Pi_2$, the procedure terminates without success.
Otherwise, refinement continues.

\paragraph{Step 3: Termination conditions.}
If every class in $\Pi_1$ and $\Pi_2$ is a singleton, the induced bijection is formed and verified explicitly on the original graphs $\mathcal{G}_1,\mathcal{G}_2$ by adjacency preservation. 
If non-singleton classes remain, the algorithm checks whether each such class is a twin class in the corresponding original graph:
a non-singleton class $\mathcal{C}$ is accepted as a true-twin class if all distinct $u,v\in \mathcal{C}$ are adjacent and satisfy $N(u)\setminus \mathcal{C} = N(v)\setminus \mathcal{C}$,
and as a false-twin class if all distinct $u,v\in \mathcal{C}$ are non-adjacent and satisfy $N(u)\setminus \mathcal{C} = N(v)\setminus \mathcal{C}$.
If, for every signature class, the twin-kind matches across $\mathcal{G}_1$ and $\mathcal{G}_2$, a canonical bijection is constructed within each class (by deterministic ordering) and verified explicitly on $\mathcal{G}_1,\mathcal{G}_2$.
If verification succeeds, the graphs are declared isomorphic; otherwise the algorithm fails. 

\paragraph{Step 4: Select an unresolved matched class.}
If termination is not reached, we select an unresolved matched signature class according to a fixed strategy. 
In one run of the algorithm, we always choose a smallest signature class $\mathcal{C}_1$ in $\Pi_1$ with $|\mathcal{C}_1|\ge 2$ whose signature type appears in $\Pi_2$ with the same size. 
In a second, independent run, we instead always choose a largest such class. 
Let $\mathcal{C}_2$ denote the corresponding class in $\Pi_2$. 
Ties are broken deterministically (e.g., lexicographically by signature). 
The next refinement is focused on $\mathcal{C}_1,\mathcal{C}_2$. 
We conclude that the inputs are isomorphic if either run of the algorithm reports isomorphism.

\paragraph{Step 5: Structured probing (temporary refinement).}
Structured probing refines $\mathcal{C}_1$ and $\mathcal{C}_2$ by computing a probe profile for each candidate vertex.
All probing is performed on temporary copies and discarded afterwards.

\subsection{Probe gadgets, probe outcomes, and probe profiles}

Fix an integer $q\ge 1$ (the current gadget size parameter), and fix two distinct nonnegative path lengths $\ell_u \neq \ell_v$.
A shared clique gadget attaches a fresh $q$-clique and connects every vertex in it to each endpoint in a given endpoint set.
A private clique gadget of size $t$ attached to $x$ adds a fresh clique of size $t$ and connects $x$ to every vertex in that clique.
A path gadget of length $\ell$ attached to $x$ adds a fresh simple path of $\ell$ new vertices with one endpoint adjacent to $x$ (and no other new edges).

\paragraph{Pair probe construction.}
Given a base graph $\widetilde{\mathcal{G}}$, distinct original vertices $u\neq v$, parameters $q$, private clique sizes $a\neq b$, and path lengths $\ell_u\neq \ell_v$, define the temporary probed graph
\[
H^{(2)}(\widetilde{\mathcal{G}};u,v;q,a,b,\ell_u,\ell_v)
\]
as $\widetilde{\mathcal{G}}$ augmented by:
(i) a shared $q$-clique connected to $\{u,v\}$,
(ii) a private clique of size $a$ attached to $u$,
(iii) a private clique of size $b$ attached to $v$,
(iv) a path of length $\ell_u$ attached to $u$ and a path of length $\ell_v$ attached to $v$ (if the length is $0$, no path is added).

Let $\Pi(H^{(2)})$ be the induced original-only signature partition and let
\[
\mathrm{End}(H^{(2)},x) := \big(\tau_x,\, s_x\big)
\]
where $\tau_x$ is the signature type of $x$ in $\Pi(H^{(2)})$ and $s_x$ is the size of its signature class.
Let
\[
\mathrm{TypeMultiset}(H^{(2)}) \quad \text{and} \quad \mathrm{SizeMultiset}(H^{(2)})
\]
denote respectively the sorted multiset of signature types of all original vertices and the sorted multiset of class sizes in $\Pi(H^{(2)})$.

The pair outcome key is then
\[
\mathrm{OK}^{(2)}(H^{(2)};u,v)
=
\Big(
\mathrm{canon}\{\mathrm{End}(H^{(2)},u),\mathrm{End}(H^{(2)},v)\},\;
\mathrm{TypeMultiset}(H^{(2)}),\;
\mathrm{SizeMultiset}(H^{(2)})
\Big),
\]
where $\mathrm{canon}\{\cdot,\cdot\}$ denotes canonical ordering of the two endpoints (so that swapping $u$ and $v$ yields the same key).
All outcome keys are computed from an original-only signature partition of the temporary probed graph.

\paragraph{Triplet probe construction.}
Given $\widetilde{\mathcal{G}}$ and distinct original vertices $u,v,w$, define the temporary probed graph
\[
H^{(3)}(\widetilde{\mathcal{G}};u,v,w;q,a,b,c,\ell_u,\ell_v,\ell_w)
\]
by attaching a shared $q$-clique to $\{u,v,w\}$, three distinct private cliques of sizes $a,b,c$ at $u,v,w$, and three paths of pairwise distinct lengths $\ell_u,\ell_v,\ell_w$ at $u,v,w$.
Let $\Pi(H^{(3)})$ be the induced original-only signature partition.
The triplet outcome key is defined by
\[
\begin{aligned}
\mathrm{OK}^{(3)}(H^{(3)};u,v,w)
&=
\Big(
\mathrm{canon}\{\mathrm{End}(H^{(3)},u),\mathrm{End}(H^{(3)},v),\mathrm{End}(H^{(3)},w)\}, \\
&\qquad
\mathrm{TypeMultiset}(H^{(3)}),\;
\mathrm{SizeMultiset}(H^{(3)})
\Big).
\end{aligned}
\]
where $\mathrm{canon}\{\cdot,\cdot,\cdot\}$ denotes canonical ordering of the three endpoints.

\paragraph{Two-orientation symmetrization.}
To make probing invariant to endpoint orientation, each probe uses two gadget assignments and records the unordered pair of resulting outcome keys.

\begin{itemize}
\item \textbf{Pairs.} Fix distinct private clique sizes $(q{+}1,q{+}2)$ and distinct path lengths $(\ell_u,\ell_v)$.
For each ordered pair $(u,v)$, compute
\[
A = \mathrm{OK}^{(2)}\!\left(H^{(2)}(\widetilde{\mathcal{G}};u,v;q,q{+}1,q{+}2,\ell_U,\ell_V);u,v\right),
\]
\[
B = \mathrm{OK}^{(2)}\!\left(H^{(2)}(\widetilde{\mathcal{G}};u,v;q,q{+}2,q{+}1,\ell_V,\ell_U);u,v\right),
\]
and let $\mathrm{POK}^{(2)}(u,v) := \mathrm{canon}\{A,B\}$ be the unordered (canonicalized) pair of keys.
\item \textbf{Triplets.} Fix private clique sizes $(q{+}1,q{+}2,q{+}3)$.
Choose a third path length $\ell_W\ge 0$ deterministically as the smallest nonnegative integer distinct from $\ell_U$ and $\ell_V$.
For each unordered $\{v,w\}$ with $v<w$ and $v,w\neq u$, compute two orientations obtained by swapping the private gadgets and path lengths assigned to $v$ and $w$, producing keys $A,B$, and define $\mathrm{POK}^{(3)}(u,v,w):=\mathrm{canon}\{A,B\}$.
\end{itemize}

\paragraph{Pair probe profile.}
For a fixed candidate vertex $u$, its pair probe profile is the multiset of symmetrized pair outcomes over all $v\neq u$:
\[
\mathrm{PP}^{(2)}_{\widetilde{\mathcal{G}}}(u;q,\ell_U,\ell_V)
\;:=\;
\left\{\!\left\{\,
\mathrm{POK}^{(2)}(u,v)
\;:\;
v\in\{0,\dots,n-1\}\setminus\{u\}
\,\right\}\!\right\},
\]
where $\{\!\{\cdot\}\!\}$ denotes a multiset.
Equivalently, we represent the profile as a frequency map
\[
\mathrm{PP}^{(2)}_{\widetilde{\mathcal{G}}}(u)
:\;
\mathrm{POK}^{(2)} \mapsto \#\{v\neq u:\mathrm{POK}^{(2)}(u,v)=\mathrm{POK}^{(2)}\}.
\]

\paragraph{Triplet probe profile.}
Analogously, the triplet probe profile is the multiset of symmetrized triplet outcomes over all unordered pairs $\{v,w\}$ with $v<w$ and $v,w\neq u$:
\[
\mathrm{PP}^{(3)}_{\widetilde{\mathcal{G}}}(u;q,\ell_U,\ell_V)
\;:=\;
\left\{\!\left\{\,
\mathrm{POK}^{(3)}(u,v,w)
\;:\;
v,w\in\{0,\dots,n-1\}\setminus\{u\},\;
v<w
\,\right\}\!\right\},
\]
or equivalently as its frequency map over distinct $\mathrm{POK}^{(3)}$ values.

\paragraph{Probe-induced refinement of a class.}
Given $\mathcal{C}\subseteq\{0,\dots,n-1\}$, probing partitions $\mathcal{C}$ into subclasses by equality of probe profiles:
\[
u \equiv v \quad\Longleftrightarrow\quad \mathrm{PP}^{(\cdot)}_{\widetilde{\mathcal{G}}}(u)=\mathrm{PP}^{(\cdot)}_{\widetilde{\mathcal{G}}}(v).
\]
Pair probing is attempted first; triplet probing is used only if pair probing yields a single group.

\subsection{Deterministic individualization by permanent augmentation}

After probing is performed on $\mathcal{C}_1$ and $\mathcal{C}_2$, refined groups are matched across graphs by exact equality of probe profiles.

Permanent augmentation then proceeds deterministically:
\begin{itemize}
\item If probing yields no singleton refined groups in either graph, choose the smallest matched refined group pair and attach a permanent clique of size $q$ to a fixed representative vertex in each graph.
\item If probing yields singleton refined groups, attach permanent cliques of size $q$ to \emph{all} matched singleton pairs (in deterministic order).
\end{itemize}
The size parameter $q$ is increased by $1$ after each matched attachment, ensuring distinctiveness of successive augmentations.
The working graphs $\widetilde{\mathcal{G}}_1,\widetilde{\mathcal{G}}_2$ are updated permanently, and the procedure proceeds to the next iteration.

\subsection{Termination and correctness}

The algorithm declares isomorphism only after constructing a bijection $f$ on original vertices and verifying adjacency preservation on the unmodified inputs $\mathcal{G}_1$ and $\mathcal{G}_2$:
\[
\forall 0\le u<v<n:
\quad
\{u,v\}\in E_1 \iff \{f(u),f(v)\}\in E_2.
\]
Hence it never declares two non-isomorphic graphs to be isomorphic, and any positive identification is certified by explicit verification on the original inputs. 
If it terminates without a verified bijection, it reports failure to resolve isomorphism under the enforced bounds and refinement regime and does not assert isomorphism.

\section{Time and space complexity}

We analyze the worst-case running time and memory usage of the implementation by decomposing it into its principal components:
(i) spectral--geometric computation,
(ii) BFS-curvature signature construction and comparison,
(iii) structured probing (triplets in the worst case),
(iv) permanent individualization. 

Let $n = |V|$ and $m = |E|$ denote the number of vertices and edges of the original input graphs.
All bounds are worst-case under the standard RAM model with arithmetic on $O(\log n)$-bit integers; spectral computations are costed using dense linear algebra. 

\paragraph{Worst-case assumptions used in this section.}
We assume:
\begin{enumerate}
\item The number of subdivision rounds is a fixed constant, hence the working graph size after subdivision remains $N = \Theta(n+m)$ up to constant factors.
In particular, when $m=\Theta(n^2)$ (dense graphs), $N=\Theta(n^2)$.
\item The algorithm performs at most $n$ rounds of permanent individualization, and hence at most $O(n)$ permanent clique attachments in the worst case. 
\item In every refinement round, pair probing fails and triplet probing is invoked.
\item Clique sizes and path lengths are initialized as constants but may increase as a function of $n$ due to repeated individualizations; we therefore track their contribution to the working graph size explicitly.
\end{enumerate}

\subsection{Working-graph size under worst-case augmentation}

Let $\widetilde{G}$ denote a working graph at some round.
Let $p$ be the number of permanent clique attachments performed so far, with $p \le n$ in the worst case.
Let $q_j$ be the clique size used in the $j$-th permanent attachment.
Since $q_j$ increases by $1$ after each attachment and begins at a constant,
\[
q_j = \Theta(j).
\]
Each attachment adds $\Theta(q_j)$ new vertices and $\Theta(q_j^2)$ new edges (the clique itself), plus $\Theta(q_j)$ incident edges to the attachment vertex.
Therefore, after $p$ attachments,
\[
N(p) \;=\; n + O(m) + \sum_{j=1}^{p} \Theta(q_j)
\;=\; n + O(m) + \Theta\!\left(\sum_{j=1}^{p} j\right)
\;=\; n + O(m) + \Theta(p^2),
\]
and
\[
M(p) \;=\; m + \sum_{j=1}^{p} \Theta(q_j^2)
\;=\; m + \Theta\!\left(\sum_{j=1}^{p} j^2\right)
\;=\; m + \Theta(p^3).
\]
In the worst case $p=\Theta(n)$, hence
\[
N(p)=\Theta(n+m+n^2)=\Theta(m+n^2),\qquad
M(p)=\Theta(m+n^3).
\]
For dense graphs ($m=\Theta(n^2)$) this yields $N=\Theta(n^2)$ and $M=\Theta(n^3)$.
For sparse graphs ($m=\Theta(n)$), this yields $N=\Theta(n^2)$ and $M=\Theta(n^3)$ as well; the permanent cliques dominate.

Finally, note that path gadgets added during probing are temporary and contribute only $O(q_j)$ vertices/edges per probe instance; this does not change asymptotic bounds because the dominant term arises from the permanent clique growth captured above.

\subsection{Cost of one spectral-geometric evaluation}

In each refinement round, the algorithm performs a full Laplacian eigendecomposition on the current working graph (for each of the two graphs).
Using dense linear algebra, for a graph with $N$ vertices this costs
\[
O(N^3) \text{ time and } O(N^2) \text{ space}.
\]
All additional spectral-dimension estimation and coefficient extraction steps are lower order in $N$ relative to the dense eigendecomposition cost $O(N^3)$. 

Thus, a single spectral-geometric evaluation stage costs $O(N^3)$ time and $O(N^2)$ space per graph, and the same asymptotic bounds apply to the two-graph execution up to a constant factor.

\subsection{Cost of BFS-curvature signature construction}

Constructing signatures for all original vertices requires performing BFS explorations on the working graph.
A BFS costs $O(N+M)$, hence performing BFS from all $n$ original vertices costs
\[
O\big(n(N+M)\big).
\]
In the worst case, this is dominated by the spectral cost $O(N^3)$ once $N=\Omega(n^2)$, which holds under the permanent clique growth.

\subsection{Worst-case cost of structured probing (triplets)}

Let $\mathcal{C}$ be the selected unresolved class with $|\mathcal{C}|=k$.
In the worst case, $k=\Theta(n)$.

Triplet probing computes a probe profile for each $u\in\mathcal{C}$ by evaluating all unordered pairs $\{v,w\}\subseteq \{0,\dots,n-1\}\setminus\{u\}$, i.e.,
\[
\binom{n-1}{2} = \Theta(n^2)
\]
probes per $u$.
Therefore, the number of probe evaluations per round is
$
\Theta(k n^2)=\Theta(n^3).
$

Each probe evaluation constructs a temporary augmented graph and recomputes:
(i) a spectral decomposition and
(ii) the induced original-only signature partition.
Both are computed on a graph whose size is $N$, hence the per-probe cost is
$
O(N^3)
$
in the dense eigendecomposition model. 
Therefore, the worst-case time for triplet probing in a single refinement round is
\[
\Theta(n^3)\cdot O(N^3) \;=\; O(n^3 N^3).
\]

\subsection{Worst-case number of rounds}

We assume at most $n$ permanent individualization steps.
Each refinement round performs at least one permanent attachment (either one matched representative pair or multiple matched singleton pairs), hence the total number of refinement rounds is at most
$
R = O(n).
$

\subsection{Total worst-case time bound}

Combining the above:
\begin{itemize}
\item per round worst-case cost is dominated by triplet probing: $O(n^3 N^3)$,
\item number of rounds is $O(n)$,
\item and in the worst case $N=\Theta(n^2)$ due to the cumulative clique growth (even if the original graph is sparse).
\end{itemize}

Thus the total worst-case time is
\[
O\big(n \cdot n^3 N^3\big) = O(n^4 N^3).
\]
Substituting $N=\Theta(n^2)$ gives the explicit worst-case bound
$
O(n^{10}).
$
This is a conservative worst-case bound for the present implementation under the assumptions stated at the beginning of the section. In particular, it reflects repeated dense spectral recomputation during probing and is not intended as a claim of practical optimality. 

Thus, while the current implementation has a high conservative worst-case bound, the overall procedure remains deterministic and polynomial in the graph size and refinement parameters.

\subsection{Worst-case space bound}

The dominant memory cost in any stage arises from storing dense spectral objects for the working graph:
the Laplacian (or an equivalent dense representation) and the eigenvector matrix.
This requires
$
O(N^2)
$
space per graph, hence still $O(N^2)$ overall up to constants.
In the worst case $N=\Theta(n^2)$, therefore the worst-case space usage is
$
O(n^4).
$

Temporary probe graphs do not asymptotically increase peak memory beyond $O(N^2)$ because they can be constructed and discarded per probe evaluation; peak usage remains dominated by the spectral matrices.

\subsection{Verification cost}

When a candidate bijection is constructed, explicit verification checks adjacency preservation on the original graphs.
This requires $O(m)$ time (or $O(n^2)$ in dense representation) and $O(1)$ additional working memory beyond the input storage, and thus does not affect the asymptotic worst-case bounds above.

\section{Implementation details}

This section records implementation-level decisions that ensure
determinism, numerical stability, and reproducibility.
Only aspects not already covered in the algorithmic description are discussed here.

\subsection*{Subdivision policy}

Subdivision depth is determined once during initialization.
The smallest subdivision level satisfying prescribed vertex-count bounds
and non-degenerate spectral-dimension estimation is selected.
Subdivision is applied symmetrically and is not increased later.

\subsection*{Spectral evaluation parameters}

Each refinement round computes:

\begin{itemize}
\item the combinatorial Laplacian,
\item a full eigendecomposition,
\item a spectral-dimension estimate,
\item a short-time curvature coefficients on a deterministically updated time interval.
\end{itemize}

The number of time samples, coefficient depth,
and regression parameters are controlled by fixed global constants.
These values remain unchanged throughout execution.

\subsection*{Coefficient quantization}

Curvature coefficients are floating-point values
derived from spectral computations.
They are discretized using deterministic scale-based rounding.

For a coefficient value $x$:

\begin{enumerate}
\item Non-finite values (NaN or $\pm\infty$) are mapped to $0$.
\item The value is multiplied by a fixed scaling constant.
\item The result is rounded to the nearest integer.
\item It is then rescaled back to floating point.
\item Optionally, the value may be snapped to a fixed bucket size.
\item Values whose magnitude falls below a fixed threshold
      are clamped to zero.
\end{enumerate}

All constants governing scale, bucket size,
and thresholding are fixed globally.
The same discretization is applied to both graphs. 
This procedure prevents spurious refinement
caused by insignificant floating-point perturbations.

\subsection*{Floating-point tolerance in spectral grouping}

During construction of the curvature time grid,
near-equal spectral quantities are merged using
a tolerance that is bounded below by the local
floating-point unit-in-the-last-place (ULP).
This prevents instability when eigenvalues are extremely close. 
ULP-based tolerance is used only in this spectral grouping step,
not in curvature coefficient quantization.

\subsection*{Signature construction}

BFS-curvature signatures are constructed exclusively
for original vertices.
Auxiliary gadget vertices influence signatures only
through graph structure;
they are never assigned or compared as signature identities.

Signature keys consist of:

\begin{itemize}
\item discretized curvature coefficient vectors,
\item BFS-layer aggregation summaries,
\item structural counts derived deterministically from the graph.
\end{itemize}

All signature components are deterministically ordered.

\subsection*{Probing configuration}

When an unresolved matched class is selected:

\begin{itemize}
\item Pair probing evaluates all ordered pairs $(u,v)$
      with $u$ in the selected class and $v$ any other original vertex.
\item Two symmetric gadget orientations are evaluated.
\item Probe profiles are constructed as frequency maps
      of canonicalized outcome keys.
\item If pair probing yields a single refined group in both graphs,
      triplet probing is invoked.
\end{itemize}

Probe configurations are enumerated in lexicographic vertex order.
No heuristic prioritization or adaptive search is used; all configurations are explored deterministically. 

\subsection*{Permanent individualization}

If probing does not produce complete separation,
a permanent clique of size $q$ is attached
to matched vertices.
The parameter $q$ begins at a fixed constant 
and increases by one after each attachment. We start with $q = 3$ (a fixed constant chosen to ensure that initial augmentations are structurally distinct from the base graph). Each clique introduces fresh vertices.
Because $q$ increases monotonically,
later attachments are structurally distinguishable
from earlier ones.

\subsection*{Twin detection}

Twin classes are detected directly on the original graphs.
For each non-singleton class:

\begin{itemize}
\item true twins are verified by mutual adjacency
      and identical external neighborhoods,
\item false twins are verified by non-adjacency
      and identical external neighborhoods.
\end{itemize}

If corresponding twin classes match in both graphs,
a canonical bijection is constructed
by deterministic ordering within each class.
The bijection is then explicitly verified.

\subsection*{Determinism and reproducibility}

All constants controlling:

\begin{itemize}
\item subdivision bounds,
\item spectral coefficient depth,
\item time-grid construction,
\item quantization scale and bucket size,
\item probing gadget sizes,
\end{itemize}

are fixed in advance. 
No randomness is used at any stage of the procedure. 
All vertex selections, probe enumerations,
class selections, and attachment orders
are performed in deterministic index order. 
Given identical inputs,
the implementation produces identical refinement sequences
and identical outcomes.

\section{Experimental results}
We evaluate the procedure on randomly generated instances and on curated benchmark families commonly used in graph isomorphism studies. Across all tested instances in this work, the procedure either recovered and explicitly verified an isomorphism on the original inputs or correctly certified non-isomorphism through signature mismatch; we did not observe unresolved cases in these experiments. These results should be viewed as empirical evidence for the present implementation rather than as a provable guarantee of universal success.

\subsection{Randomly generated instances}
\begin{figure}[H]
    \centering
    \includegraphics[scale=0.7]{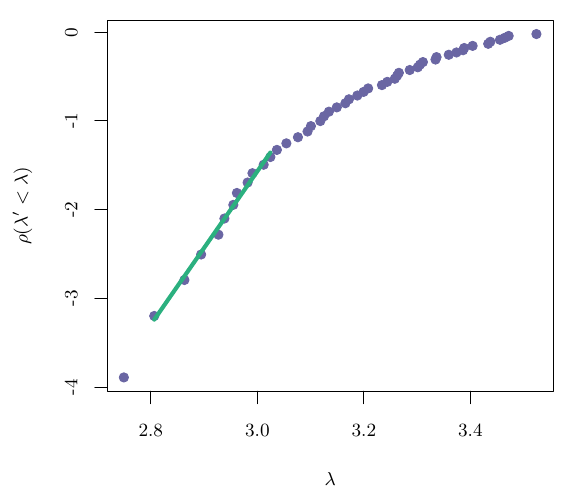}
    \caption{The spectral dimension $d_s = 17$ recovered for $\mathcal{G}_1$ with $n = 50$.}
    \label{fig:iso10}
\end{figure}

\begin{figure}[H] %
  \begin{subfigure}[c]{0.45\textwidth}
        \centering
    \includegraphics[scale=0.16]{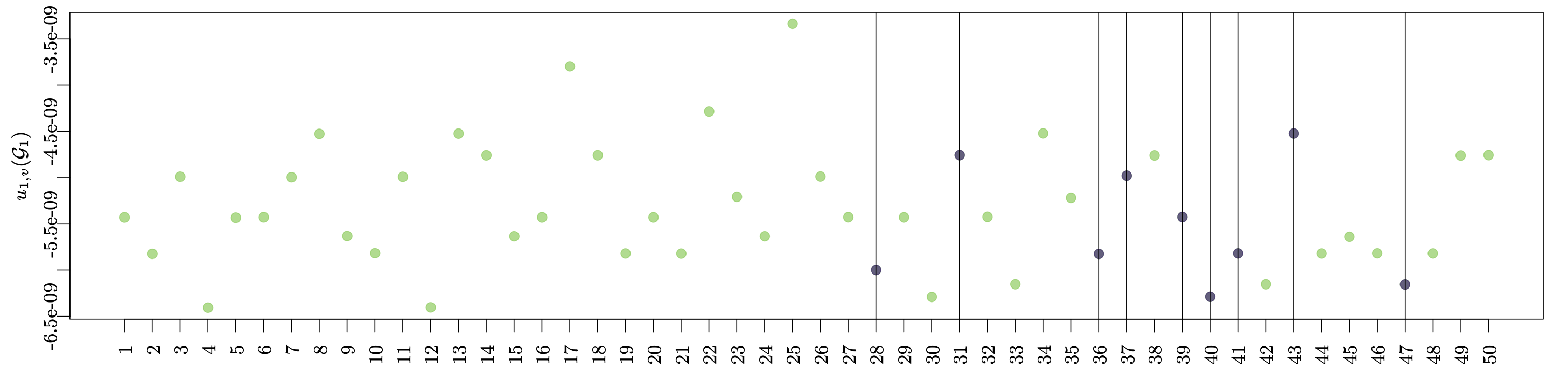}
    \caption{}
     \end{subfigure}
          \hfill

      \begin{subfigure}[c]{0.45\textwidth}
        \centering
    \includegraphics[scale=0.16]{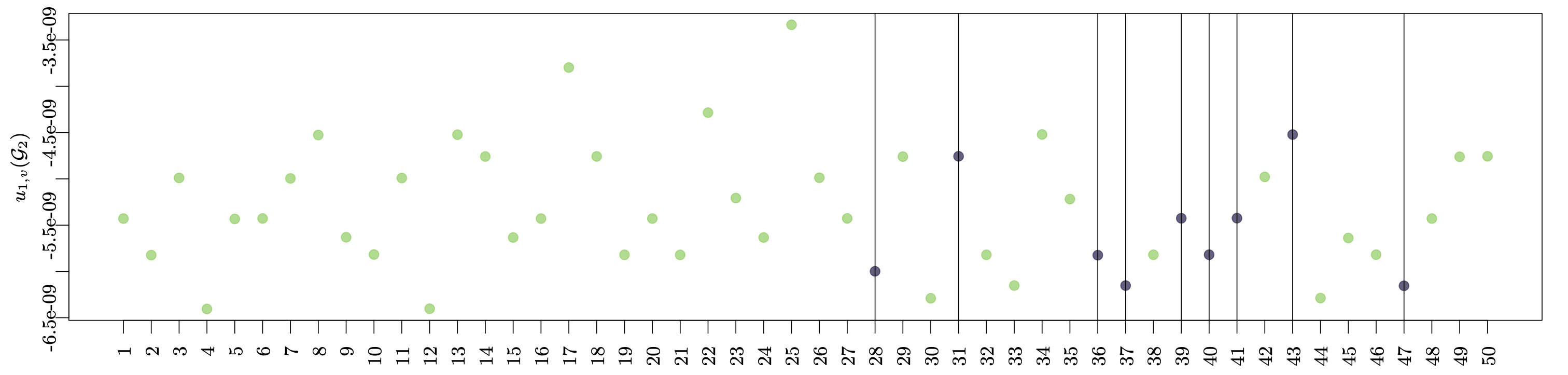}
      \caption{}
    \end{subfigure}
    \caption{The curvature-derived values $u_1(v)$ for $\mathcal{G}_1$ and $\mathcal{G}_2$. The vertical lines indicate indices of relabeled vertices. }
    \label{fig:is10curv}
\end{figure}

We begin with a representative positive instance consisting of two graphs $\mathcal{G}_1$ and $\mathcal{G}_2$ on $n=50$ vertices, where $\mathcal{G}_2$ is obtained from $\mathcal{G}_1$ by relabeling a subset of vertices (in this example, $9$ vertices are moved to different indices).
Figure~\ref{fig:iso10} shows the cumulative eigenvalue counting function $\rho(\lambda' < \lambda)$ used to estimate the spectral dimension $d_s$ on a logarithmic scale.
We then compute the leading curvature-like coefficient $u_1(v)$ for every vertex $v$ in both graphs.
Although $\mathcal{G}_1$ and $\mathcal{G}_2$ are isomorphic, the vertex ordering differs, and therefore the sequences $\{u_1(v)\}_{v \in V(\mathcal{G}_1)}$ and $\{u_1(v)\}_{v \in V(\mathcal{G}_2)}$ need not match entry-wise; this mismatch is highlighted in Figure~\ref{fig:is10curv}.

\begin{figure}[H]
    \centering
    \includegraphics{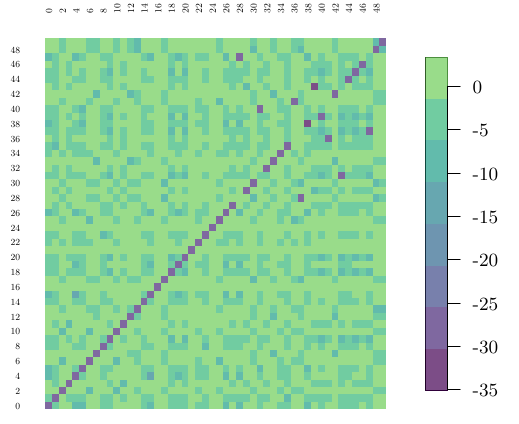}
    \caption{Heatmap of $C_{ij}$ for the isomorphic pair. The $9$ relabeled vertices appear as off-diagonal structure.}
    \label{fig:iso10heatmappingfig:gap10}
\end{figure}

\begin{figure}[H]
  \begin{subfigure}[c]{0.45\textwidth}
    \centering
    \includegraphics[scale=0.7]{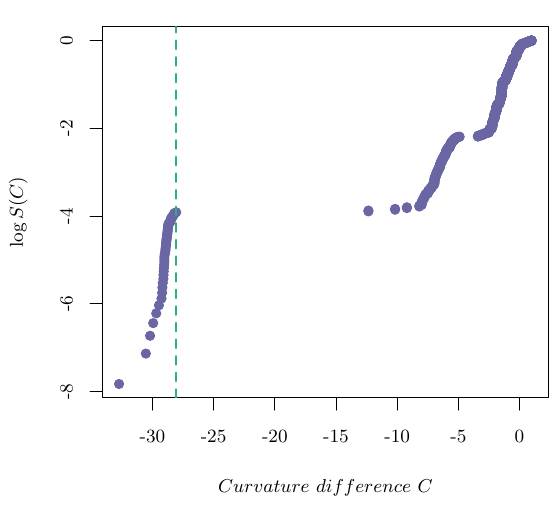}
    \caption{}
  \end{subfigure}
  \hfill
  \begin{subfigure}[c]{0.45\textwidth}
    \centering
    \includegraphics{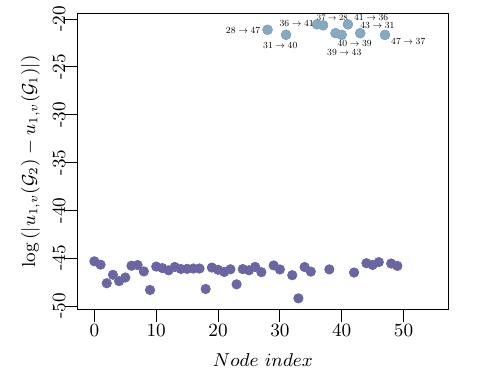}
    \caption{}
  \end{subfigure}
  \caption{(a) The number of pairs below the dashed vertical line (denoting $C_{\mathrm{critical}}$) equals $n$, consistent with isomorphism. (b) The induced rank-based pairing from sorted curvature-derived values.}
  \label{fig:gap10}
\end{figure}

To infer a correspondence, we form the pairwise difference matrix
\[
C_{ij} = \log\!\left|u_1(v_i \in V(\mathcal{G}_1)) - u_1(w_j \in V(\mathcal{G}_2))\right|,
\]
shown in Figure~\ref{fig:iso10heatmappingfig:gap10}.
Flattening and sorting the entries of $C$ yields a one-dimensional list in which a pronounced gap separates near-matching curvature pairs from the remaining pairs.
We detect the largest gap to obtain the threshold $C_{\mathrm{critical}}$ and examine the cumulative count $S(C)$; Figure~\ref{fig:gap10}(a) shows that the number of pairs below $C_{\mathrm{critical}}$ equals $n$ for this instance, consistent with isomorphism.
A vertex pairing is then obtained by sorting vertices by their curvature-derived values and matching ranks, as illustrated in Figure~\ref{fig:gap10}(b).
Finally, as in all experiments, any proposed correspondence is explicitly verified to preserve adjacency on the original graphs.

\begin{figure}[H] %
  \begin{subfigure}[c]{0.45\textwidth}
    \centering
    \includegraphics[scale=0.15]{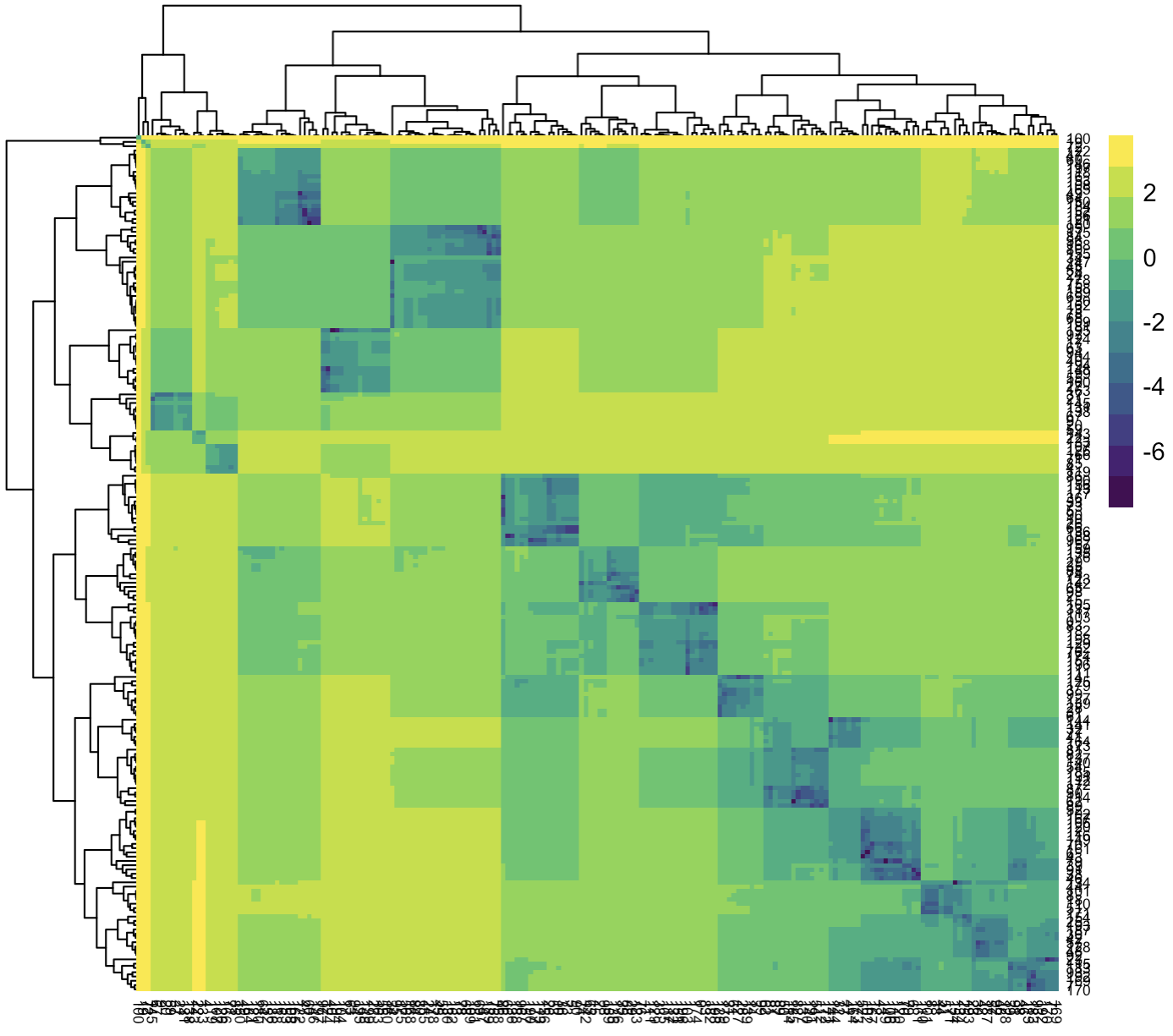}
    \caption{}
  \end{subfigure}
  \hfill
  \begin{subfigure}[c]{0.45\textwidth}
    \centering
    \includegraphics[scale=0.7]{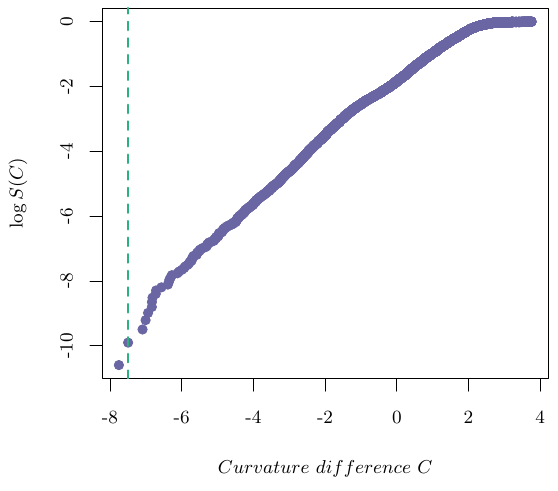}
    \caption{}
  \end{subfigure}
  \caption{Two graphs $\mathcal{G}_1$ and $\mathcal{G}_2$ are one edge-switch away from isomorphism. (a) The number of pairs below the first gap in $C_{\mathrm{critical}}$ is not $n$, and the graphs are declared non-isomorphic. (b) Heatmap of the pairwise differences $|u_1(v_i) - u_1(w_j)|$ for $n=200$ and non-homogeneous degree.}
\label{fig:si15}
\end{figure}

We also evaluate negative instances in which the two graphs are non-isomorphic.
As an illustrative example, we consider two graphs that differ by a single edge switch.
Figure~\ref{fig:si15}(a) shows the corresponding sorted curvature-difference behavior.
In contrast to the isomorphic case, the curve $\log(S(C))$ does not exhibit a clear separation into $n$ near-matching pairs; correspondingly, the number of pairs below the first prominent gap is $\ll n$, and the graphs are declared non-isomorphic.
Figure~\ref{fig:si15}(b) shows the heatmap of pairwise curvature differences for this case.

We further carry out the computation across nine graph families, generating five instances per family.
Our framework successfully resolved all instances tested in these experiments; for brevity, we show one representative pair from each class below. 

\subsubsection{Random geometric graphs}
\begin{figure}[H]
    \begin{subfigure}[c]{0.45\textwidth}
    \centering
    \includegraphics[scale=0.8]{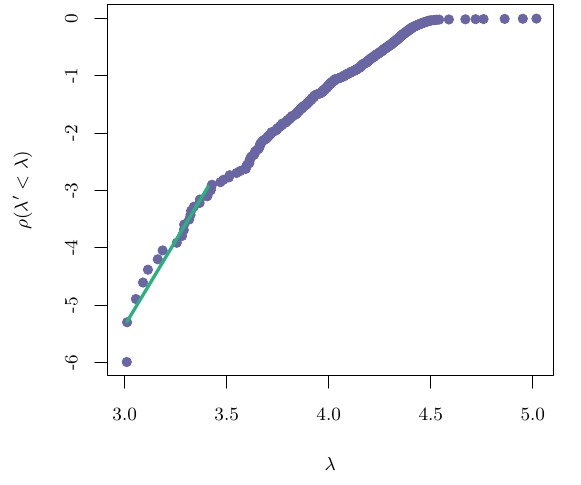}
    \caption{}
    \end{subfigure}
\hfill
    \begin{subfigure}[c]{0.45\textwidth}
    \centering
    \includegraphics[scale=0.6]{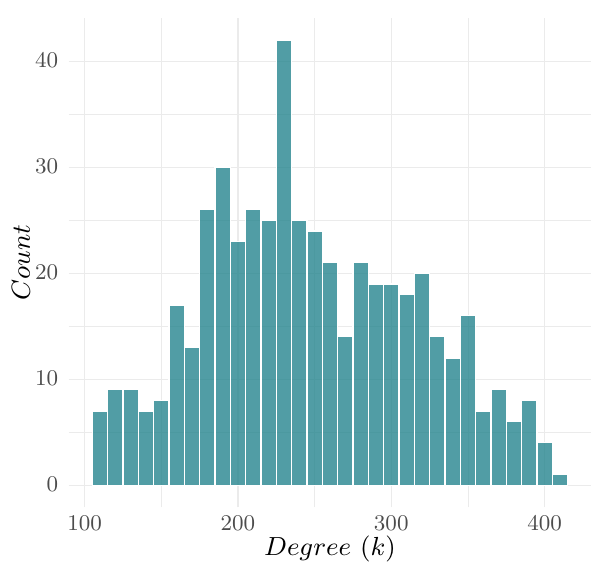}
    \caption{}
    \end{subfigure}
\hfill
    \begin{subfigure}[c]{0.45\textwidth}
    \centering
    \includegraphics[scale=0.15]{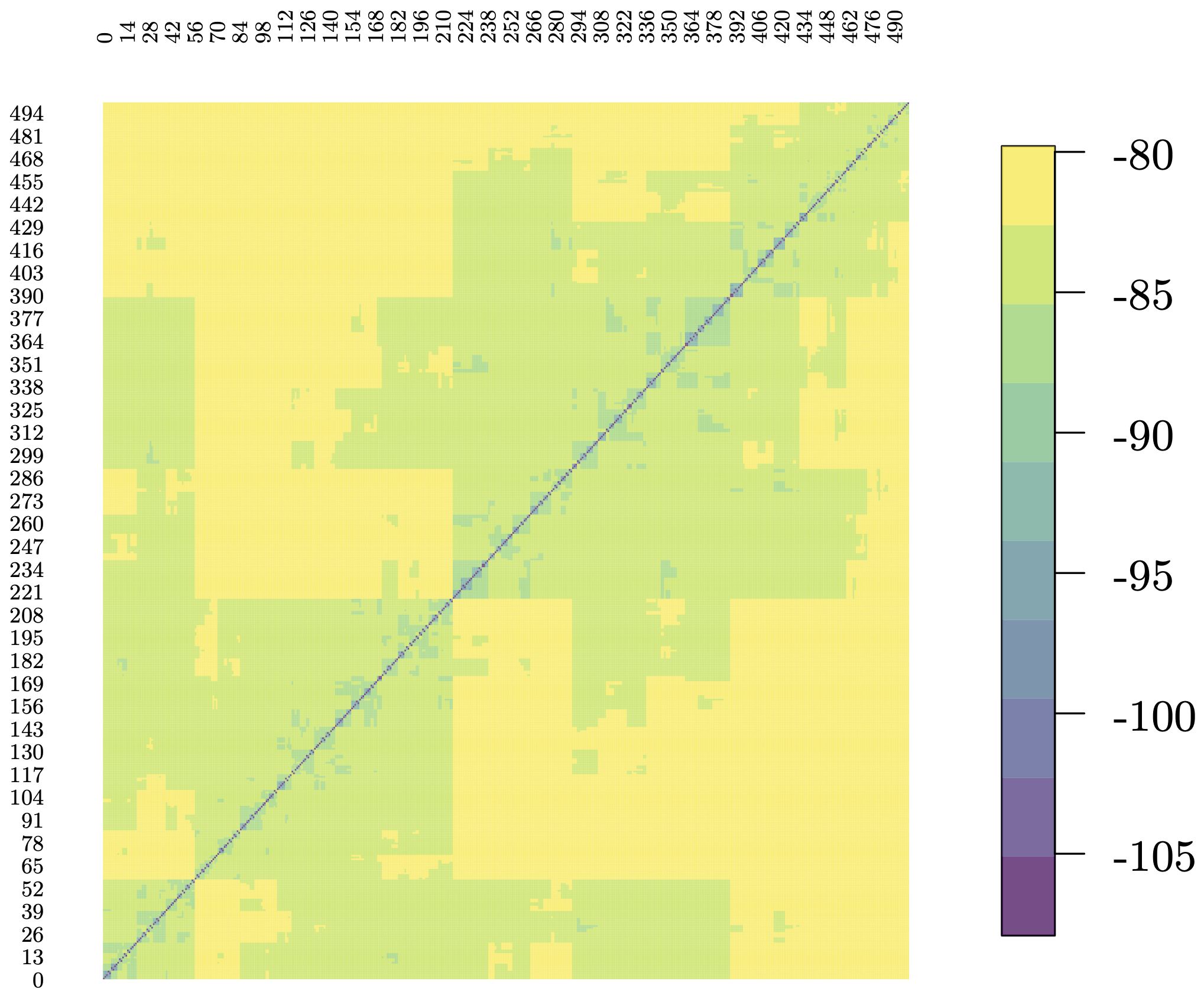}
    \caption{}
    \end{subfigure}
\caption{Recovered spectral dimension $d_s = 30.32$, degree distribution of $\mathcal{G}_1$, and heatmap of the pairwise logarithmic difference $\log|u_1(v_i)-u_1(w_j)|$ for random geometric graphs with $n = 500$.}
\label{fig:randgeom500}
\end{figure}

We apply our framework to random geometric graphs on $n = 500$ vertices with non-homogeneous degree distributions.
The eigenvalue counting function $\rho$, the degree distribution, and the heatmap of $C$ are shown in Figure~\ref{fig:randgeom500}.

\subsubsection{Power-law cluster graphs}
\begin{figure}[H]
    \begin{subfigure}[c]{0.45\textwidth}
    \centering
    \includegraphics[scale=0.8]{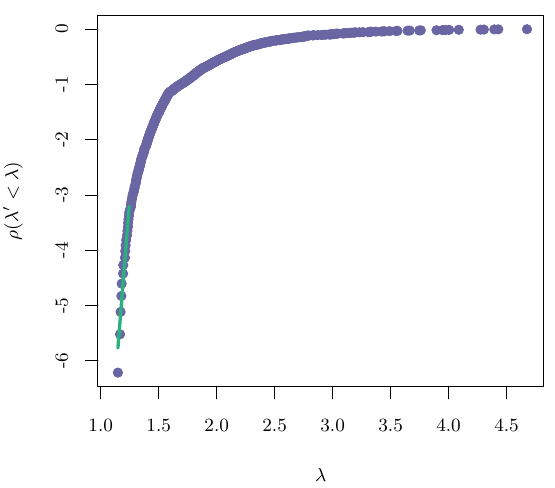}
    \caption{}
    \end{subfigure}
\hfill
    \begin{subfigure}[c]{0.45\textwidth}
    \centering
    \includegraphics[scale=0.8]{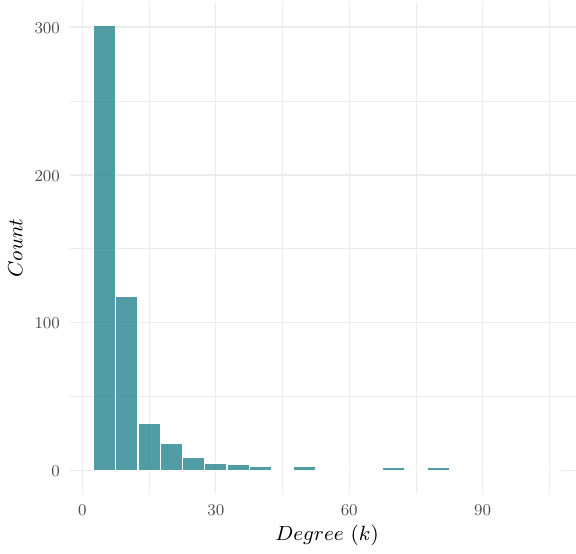}
    \caption{}
    \end{subfigure}
\hfill
    \begin{subfigure}[c]{0.45\textwidth}
    \centering
    \includegraphics[scale=0.16]{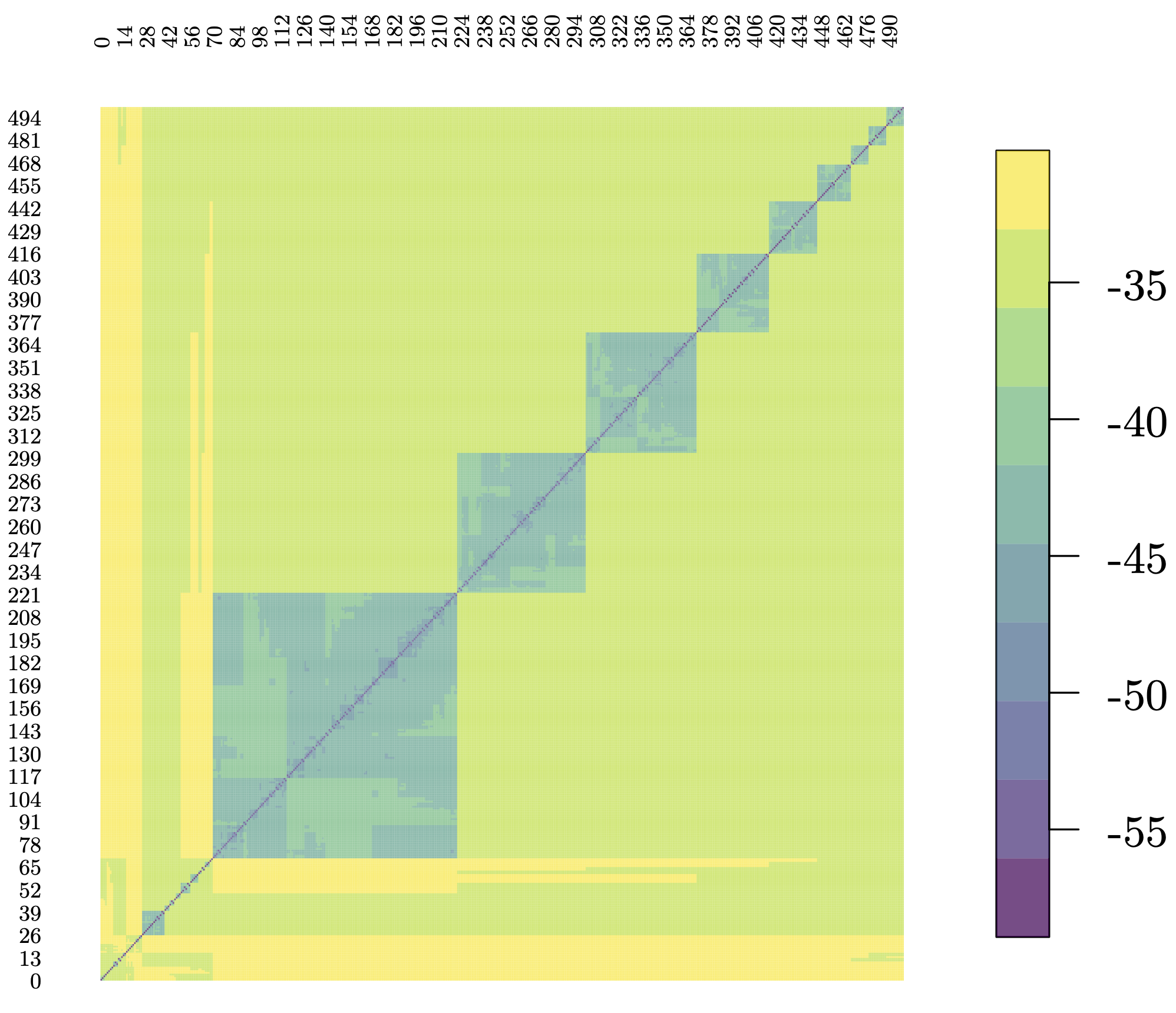}
    \caption{}
    \end{subfigure}
\caption{Recovered spectral dimension $d_s = 46.89$, degree distribution of $\mathcal{G}_1$, and heatmap of the pairwise logarithmic difference $\log|u_1(v_i)-u_1(w_j)|$ for power-law cluster graphs with $n = 500$.}
\label{fig:powerClus500}
\end{figure}

We apply our framework to power-law cluster graphs on $n = 500$ vertices with non-homogeneous degree distributions.
The eigenvalue counting function $\rho$, the degree distribution, and the heatmap of $C$ are shown in Figure~\ref{fig:powerClus500}.

\subsubsection{Stochastic block graphs}
\begin{figure}[H]
    \begin{subfigure}[c]{0.45\textwidth}
    \centering
    \includegraphics[scale=0.8]{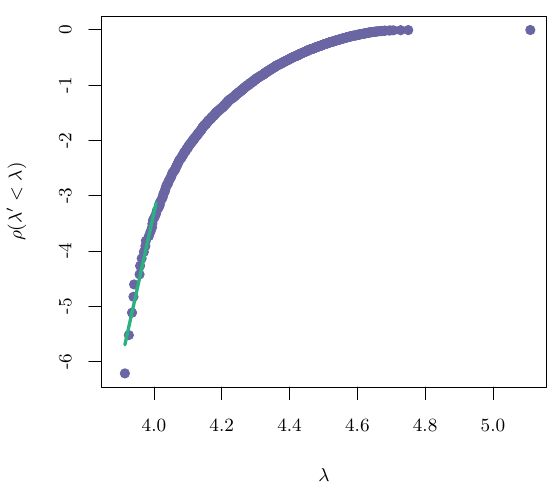}
    \caption{}
    \end{subfigure}
\hfill
    \begin{subfigure}[c]{0.45\textwidth}
    \centering
    \includegraphics[scale=0.8]{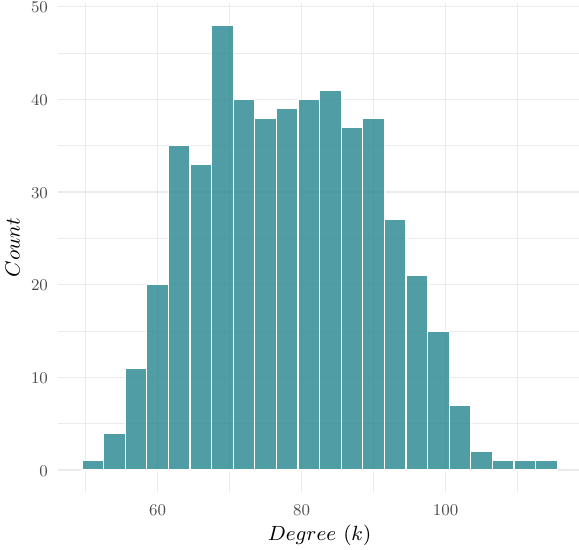}
    \caption{}
    \end{subfigure}
\hfill
    \begin{subfigure}[c]{0.45\textwidth}
    \centering
    \includegraphics[scale=0.16]{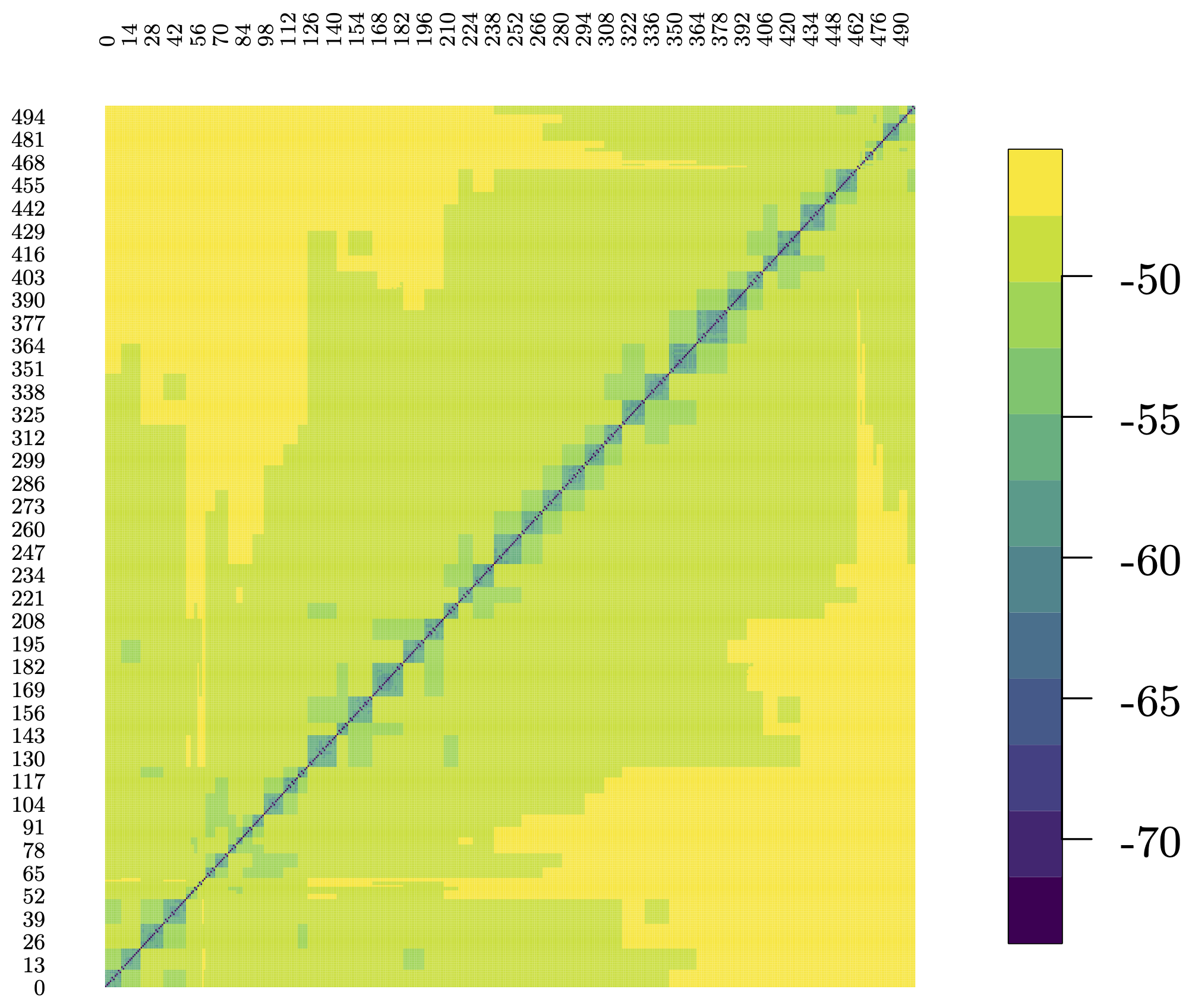}
    \caption{}
    \end{subfigure}
\caption{Recovered spectral dimension $d_s = 49$, degree distribution of $\mathcal{G}_1$, and heatmap of the pairwise logarithmic difference $\log|u_1(v_i)-u_1(w_j)|$ for stochastic block graphs with $n = 500$, with inter-community probability $p_{\mathrm{inter}}=0.1$ and intra-community probability $p_{\mathrm{intra}}=0.25$.}
\label{fig:stochasticBlock}
\end{figure}

We apply our framework to stochastic block graphs on $n = 500$ vertices with non-homogeneous degree distributions.
The eigenvalue counting function $\rho$, the degree distribution, and the heatmap of $C$ are shown in Figure~\ref{fig:stochasticBlock}.

\subsubsection{Random regular graphs}
\begin{figure}[H]
    \begin{subfigure}[c]{0.4\textwidth}
    \centering
    \includegraphics[scale=0.8]{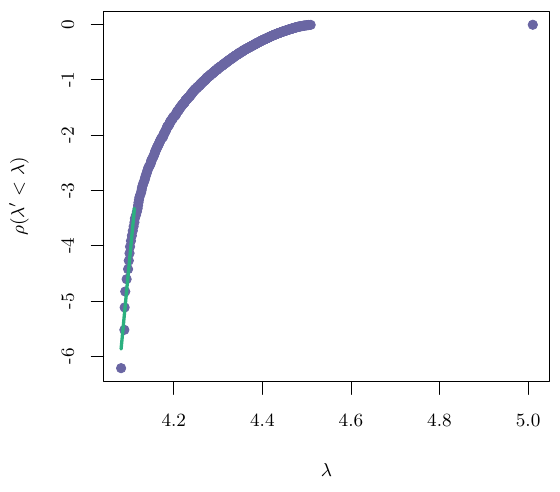}
    \caption{}
    \end{subfigure}
    \begin{subfigure}[c]{0.45\textwidth}
    \centering
    \includegraphics[scale=0.14]{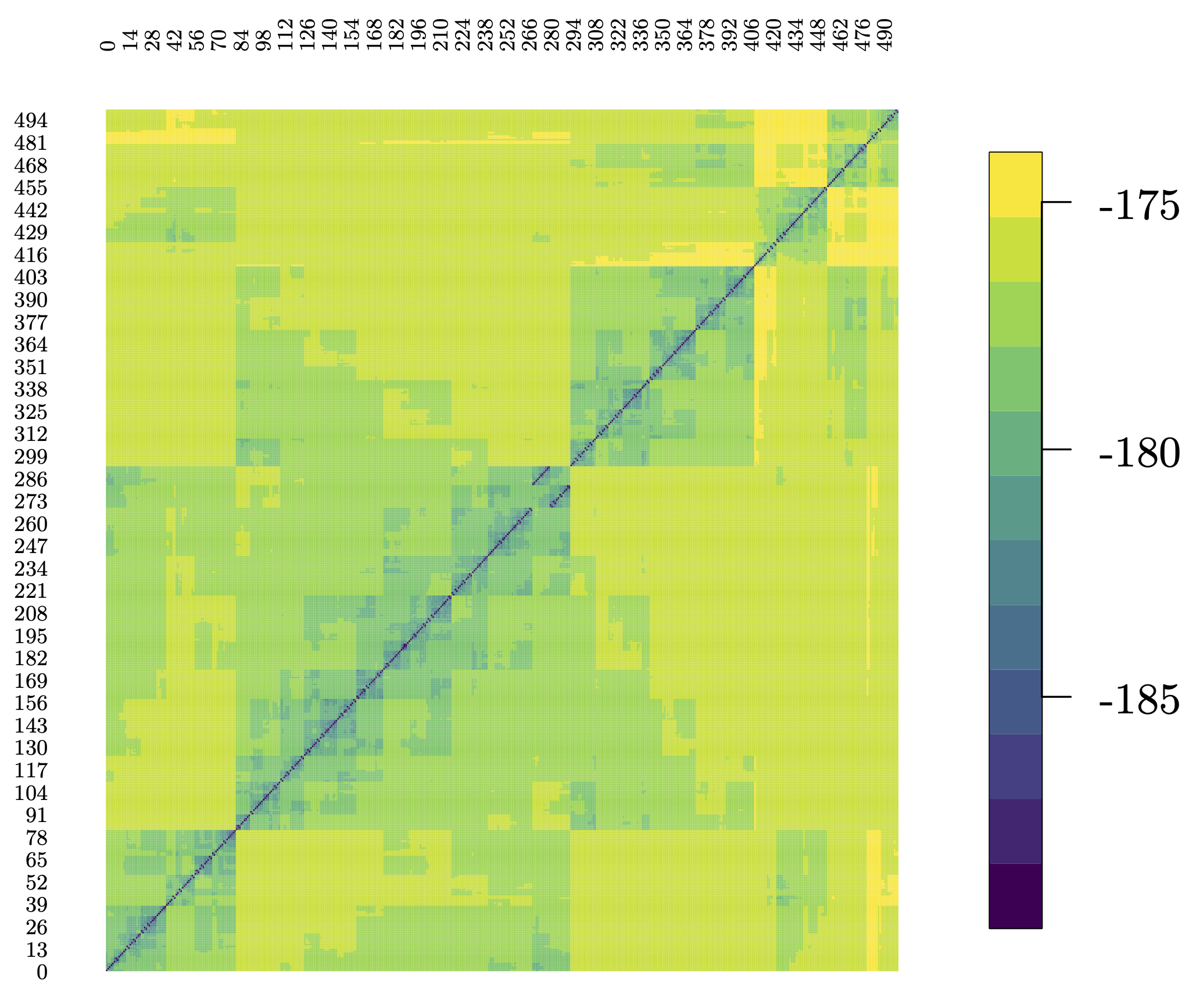}
    \caption{}
    \end{subfigure}
\caption{Recovered spectral dimension $d_s = 146.06$ and heatmap of the pairwise logarithmic difference $\log|u_1(v_i)-u_1(w_j)|$ for regular graphs with $n = 500$ and homogeneous degree $k=10$.}
\label{fig:regular}
\end{figure}

We apply our framework to random regular graphs on $n = 500$ vertices.
The eigenvalue counting function $\rho$ and the heatmap of $C$ are shown in Figure~\ref{fig:regular}.

\subsubsection{Albert--Barabási graphs}
\begin{figure}[H] %
    \begin{subfigure}[c]{0.45\textwidth}
    \centering
    \includegraphics[scale=0.8]{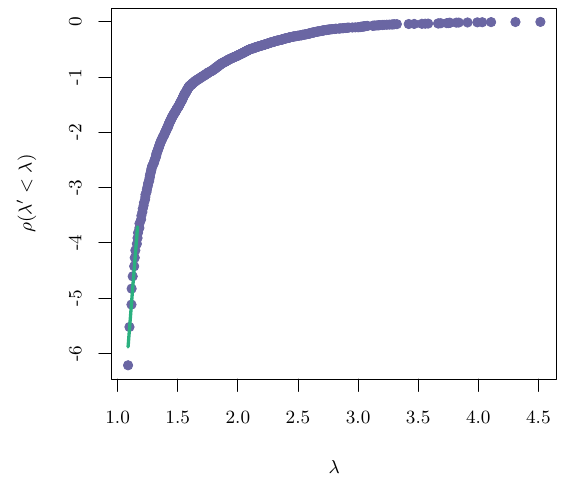}
    \caption{}
    \end{subfigure}
\hfill
    \begin{subfigure}[c]{0.45\textwidth}
    \centering
    \includegraphics[scale=0.8]{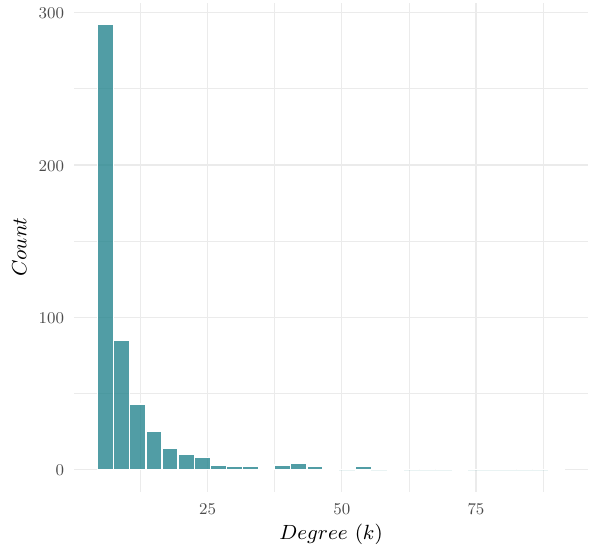}
    \caption{}
    \end{subfigure}
\hfill
    \begin{subfigure}[c]{0.45\textwidth}
    \centering
    \includegraphics[scale=0.18]{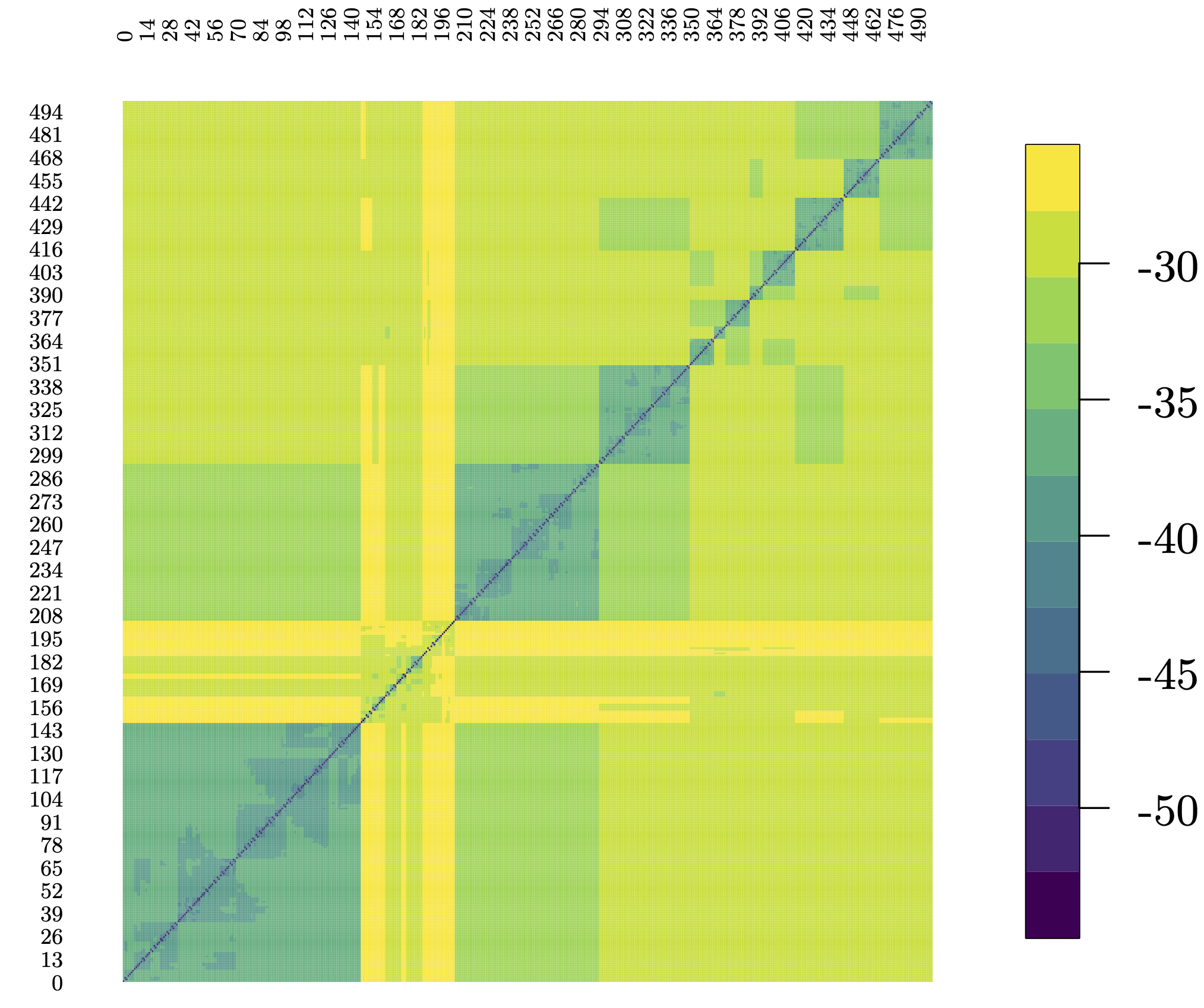}
    \caption{}
    \end{subfigure}
\caption{Recovered spectral dimension $d_s = 45.70$, degree distribution of $\mathcal{G}_1$, and heatmap of the pairwise logarithmic difference $\log|u_1(v_i)-u_1(w_j)|$ for Albert--Barabási graphs with $n = 500$, and power-law coefficient of 2.72.}
\label{fig:albertbarab}
\end{figure}

We apply our framework to Albert--Barabási graphs on $n = 500$ vertices with non-homogeneous degree distributions.
The eigenvalue counting function $\rho$, the degree distribution, and the heatmap of $C$ are shown in Figure~\ref{fig:albertbarab}.

\subsubsection{Watts--Strogatz graphs}
\begin{figure}[H] %
    \begin{subfigure}[c]{0.45\textwidth}
    \centering
    \includegraphics[scale=0.8]{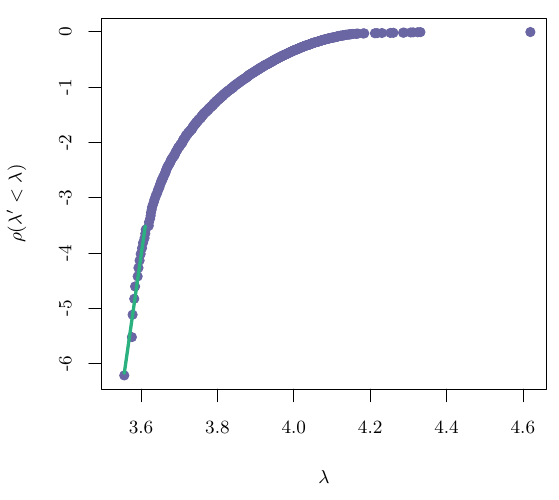}
    \caption{}
    \end{subfigure}
\hfill
    \begin{subfigure}[c]{0.45\textwidth}
    \centering
    \includegraphics[scale=0.8]{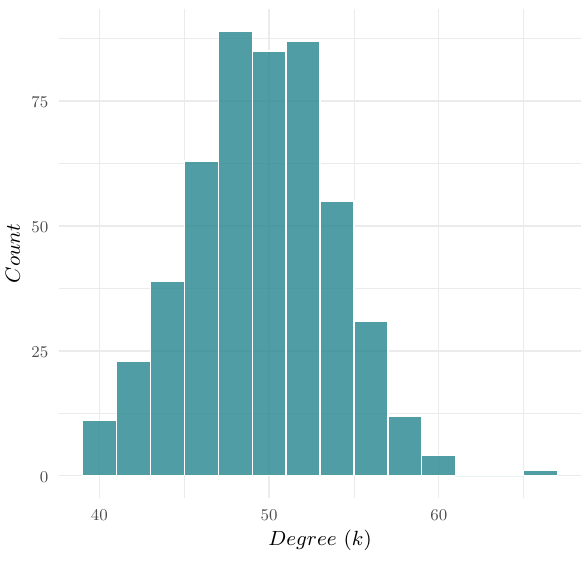}
    \caption{}
    \end{subfigure}
\hfill
    \begin{subfigure}[c]{0.45\textwidth}
    \centering
    \includegraphics[scale=0.16]{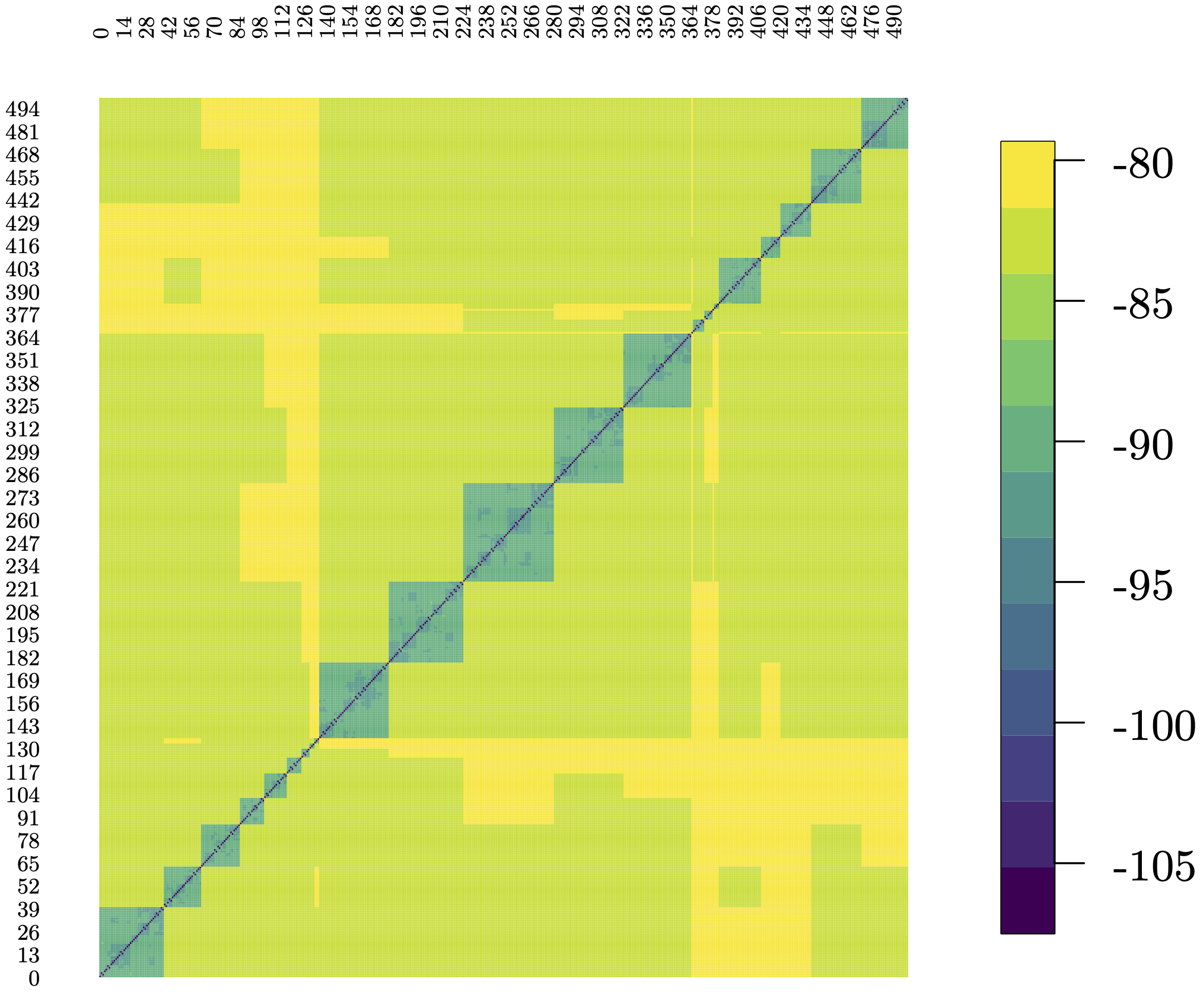}
    \caption{}
    \end{subfigure}
\caption{Recovered spectral dimension $d_s = 93.13$, degree distribution of $\mathcal{G}_1$, and heatmap of the pairwise logarithmic difference $\log|u_1(v_i)-u_1(w_j)|$ for Watts--Strogatz graphs with $n = 500$, mean degree $k_{\mathrm{mean}}=50$, and rewiring probability $0.5$.}
\label{fig:wattstr}
\end{figure}

We apply our framework to Watts--Strogatz graphs on $n = 500$ vertices with non-homogeneous degree distributions.
The eigenvalue counting function $\rho$, the degree distribution, and the heatmap of $C$ are shown in Figure~\ref{fig:wattstr}.

\subsubsection{Erd\H{o}s--R\'enyi graphs}
\begin{figure}[H] %
    \begin{subfigure}[c]{0.45\textwidth}
    \centering
    \includegraphics[scale=0.8]{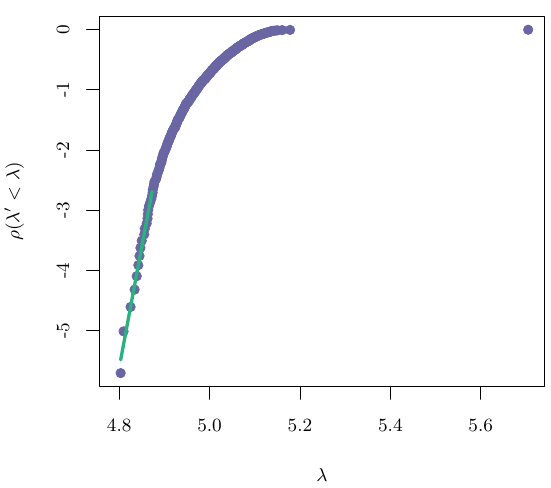}
    \caption{}
    \end{subfigure}
\hfill
    \begin{subfigure}[c]{0.45\textwidth}
    \centering
    \includegraphics[scale=0.8]{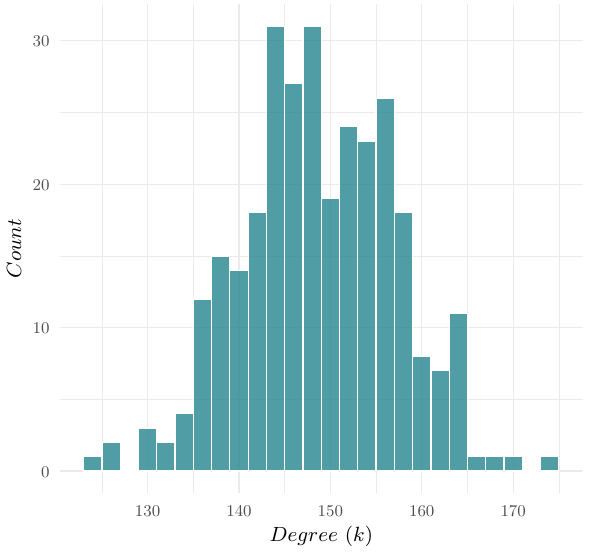}
    \caption{}
    \end{subfigure}
\hfill
    \begin{subfigure}[c]{0.45\textwidth}
    \centering
    \includegraphics[scale=0.16]{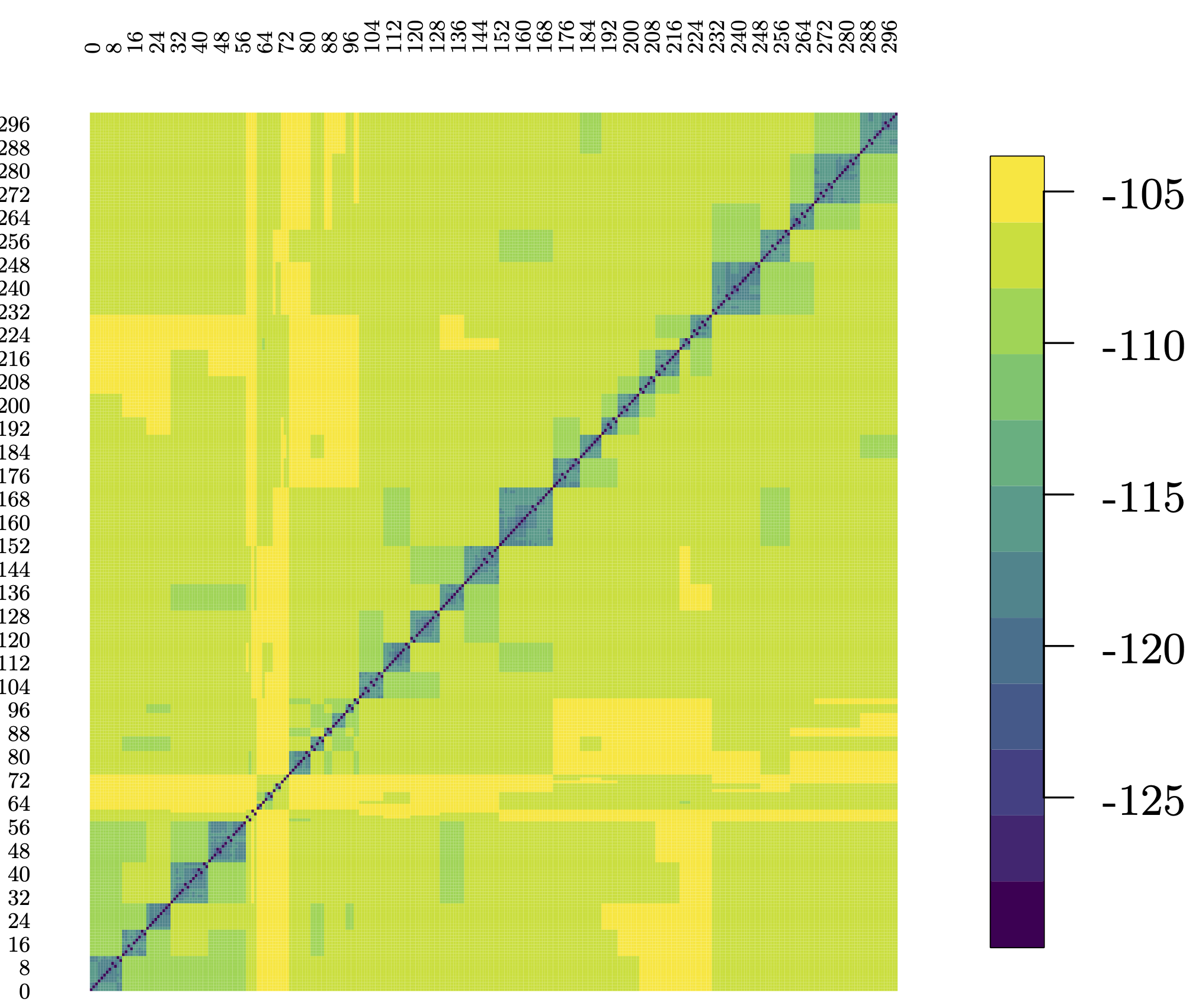}
    \caption{}
    \end{subfigure}
\caption{Recovered spectral dimension $d_s = 75.78$, degree distribution of $\mathcal{G}_1$, and heatmap of the pairwise logarithmic difference $\log|u_1(v_i)-u_1(w_j)|$ for Erd\H{o}s--R\'enyi graphs with $n = 500$ and an edge probability of $0.5$.}
\label{fig:erdren}
\end{figure}

We apply our framework to Erd\H{o}s--R\'enyi graphs on $n = 500$ vertices with non-homogeneous degree distributions.
The eigenvalue counting function $\rho$, the degree distribution, and the heatmap of $C$ are shown in Figure~\ref{fig:erdren}.

\subsubsection{Random tree graphs}
\begin{figure}[H] %
    \begin{subfigure}[c]{0.45\textwidth}
    \centering
    \includegraphics[scale=0.8]{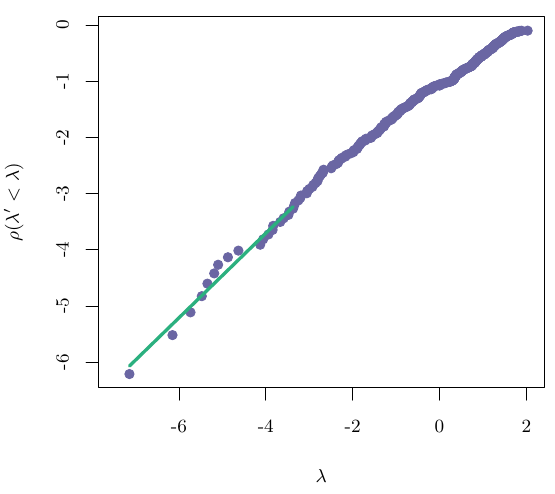}
    \caption{}
    \end{subfigure}
\hfill
    \begin{subfigure}[c]{0.45\textwidth}
    \centering
    \includegraphics[scale=0.8]{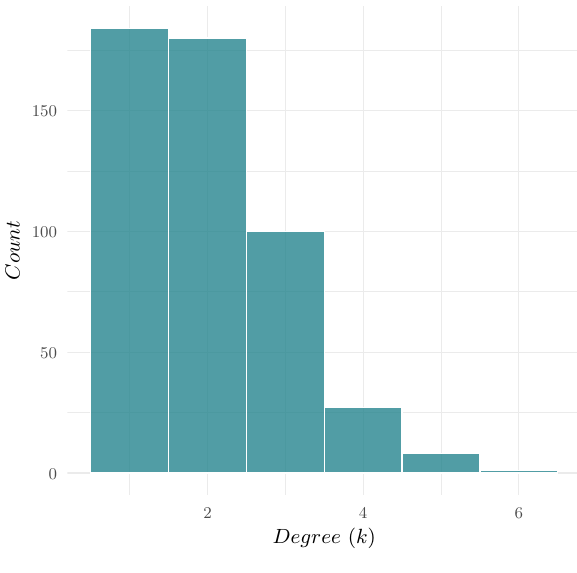}
    \caption{}
    \end{subfigure}
\hfill
    \begin{subfigure}[c]{0.45\textwidth}
    \centering
    \includegraphics[scale=0.16]{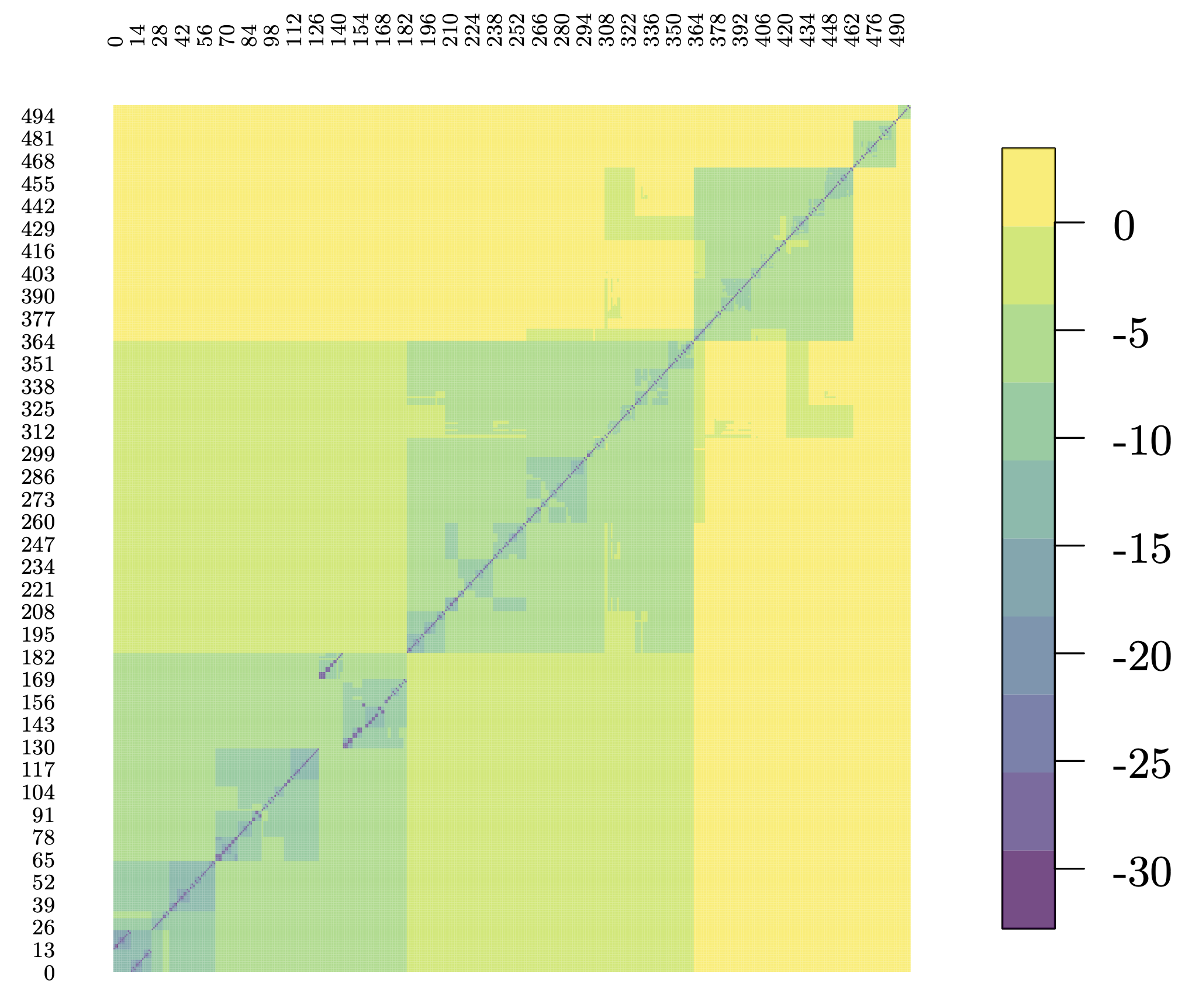}
    \caption{}
    \end{subfigure}
\caption{Recovered spectral dimension $d_s = 1.45$, degree distribution of $\mathcal{G}_1$, and heatmap of the pairwise logarithmic difference $\log|u_1(v_i)-u_1(w_j)|$ for trees with $n = 500$.}
\label{fig:trees}
\end{figure}

We apply our framework to random tree graphs on $n = 500$ vertices.
The eigenvalue counting function $\rho$, the degree distribution, and the heatmap of $C$ are shown in Figure~\ref{fig:trees}.

\subsubsection{Paley graphs}
\begin{figure}[H] %
    \begin{subfigure}[c]{0.45\textwidth}
    \centering
    \includegraphics[scale=0.14]{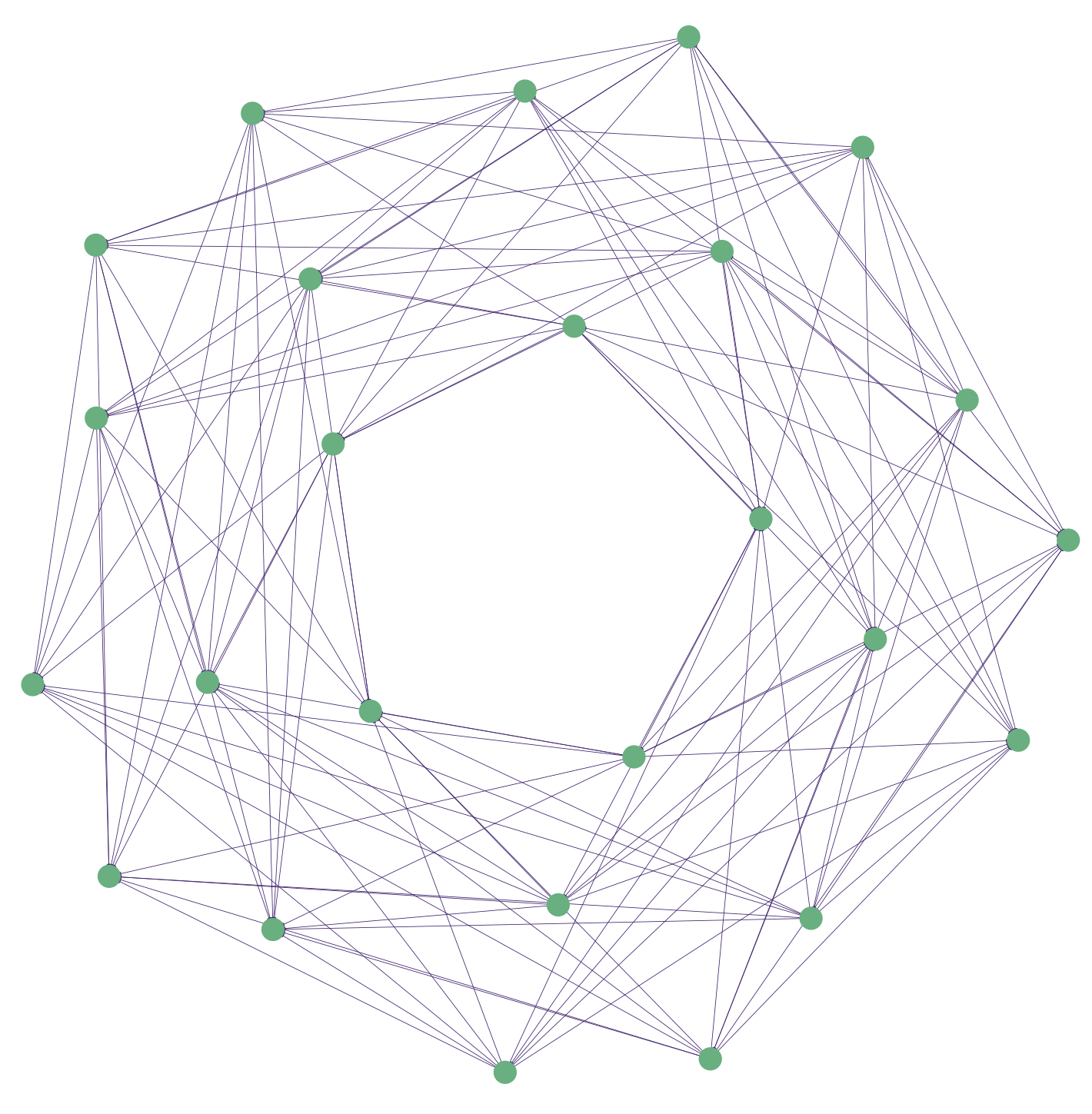}
    \caption{}
    \end{subfigure}
\hfill
    \begin{subfigure}[c]{0.45\textwidth}
    \centering
    \includegraphics[scale=0.14]{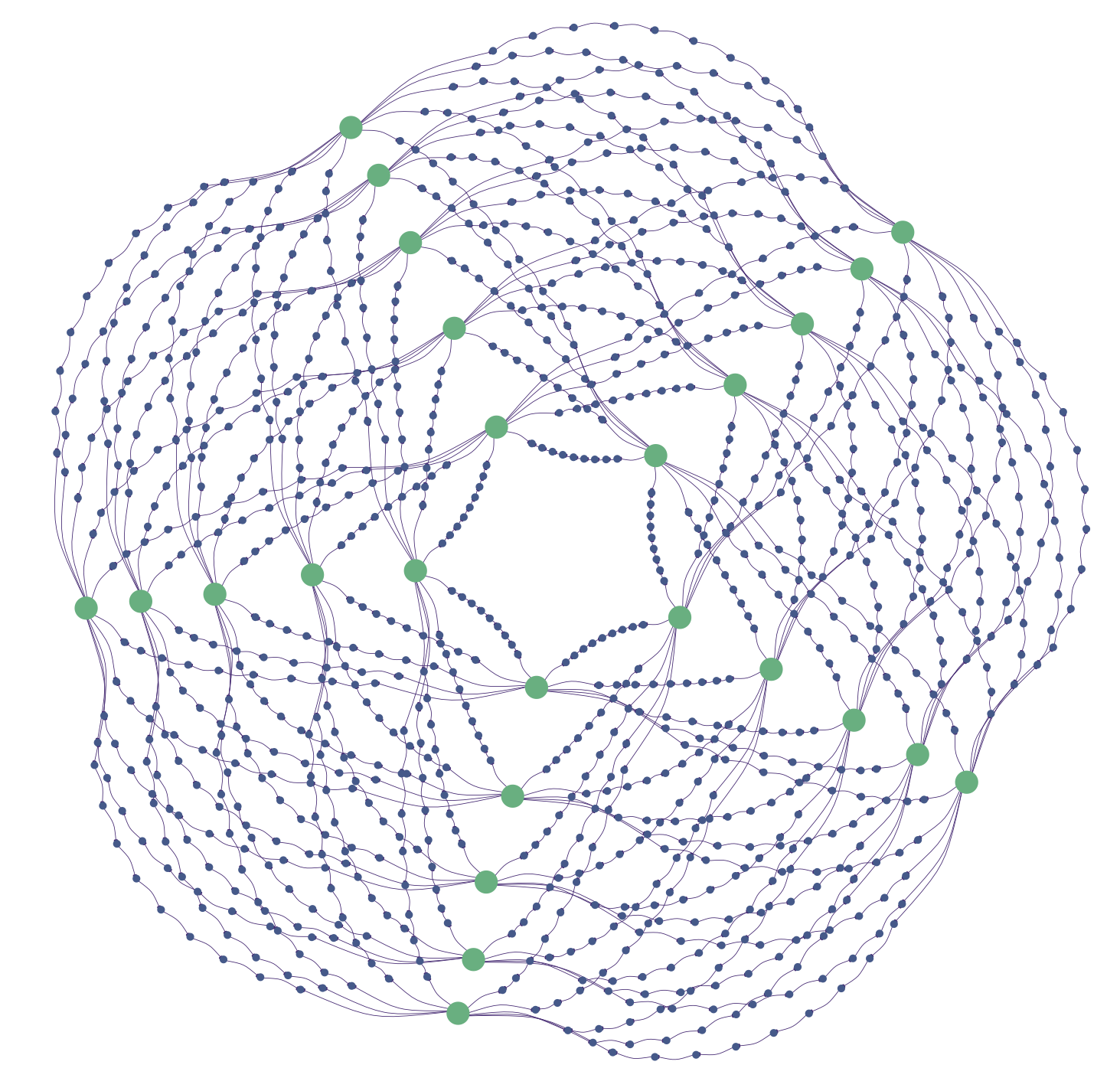}
    \caption{}
    \end{subfigure}
\caption{(a) The original Paley graph. (b) The subdivided version that ensures a power-law behavior of the eigenvalue density $\rho(\lambda'<\lambda)$.}
\label{fig:paynoblowup}
\end{figure}

\begin{figure}[H] %
    \begin{subfigure}[c]{0.45\textwidth}
    \centering
    \includegraphics[scale=0.8]{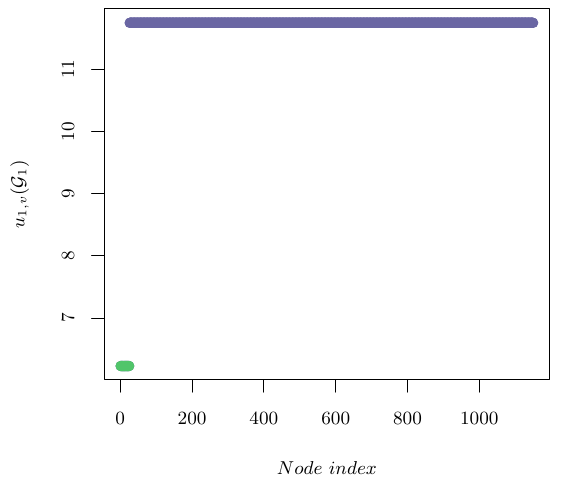}
    \caption{}
    \end{subfigure}
\hfill
    \begin{subfigure}[c]{0.45\textwidth}
    \centering
    \includegraphics[scale=0.8]{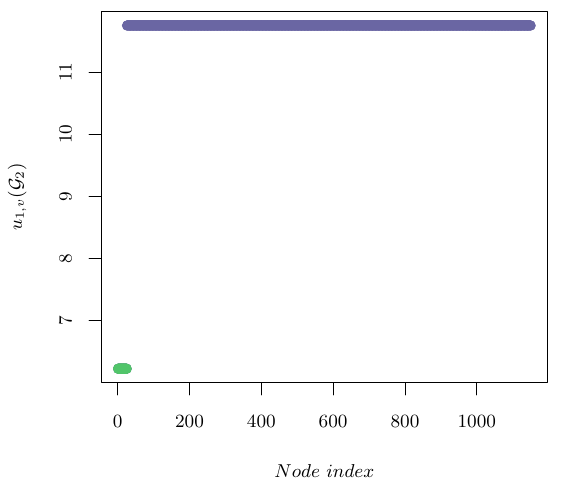}
    \caption{}
    \end{subfigure}
\caption{Curvatures $u_1(v)$ for the subdivided Paley graphs, showing alignment among the original vertices and among the auxiliary vertices (cf.\ Figure~\ref{fig:paynoblowup}(b)).}
\label{fig:paynoblowupcur}
\end{figure}

We apply our framework to Paley graphs, for which we show the non-subdivided and subdivided versions in Figure~\ref{fig:paynoblowup}.
We start by evaluating curvature on the subdivided graphs shown in Figure~\ref{fig:paynoblowupcur}.
They exhibit complete degeneracy in curvature among the original vertices, and likewise among the subdivision vertices (not all subdivision vertices share similar curvature but those differences are not visually detectable).

\begin{figure}[H] %
    \centering
    \includegraphics[scale=0.18]{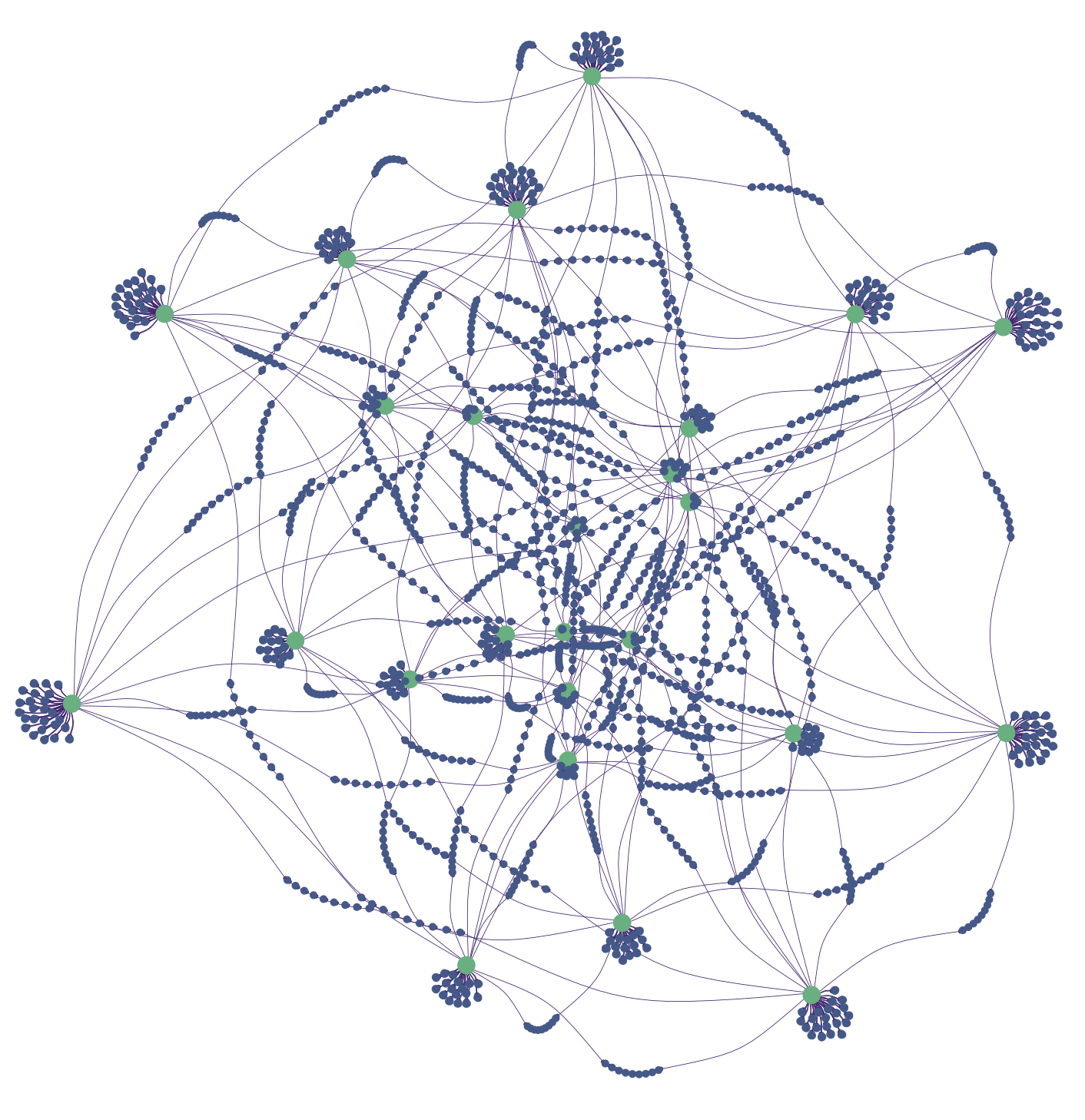}
    \caption{The resulting graph after the vertex identification/individualization construction.} 
    \label{fig:payblowup}
\end{figure}

\begin{figure}[H] %
    \begin{subfigure}[c]{0.45\textwidth}
    \centering
    \includegraphics[scale=0.8]{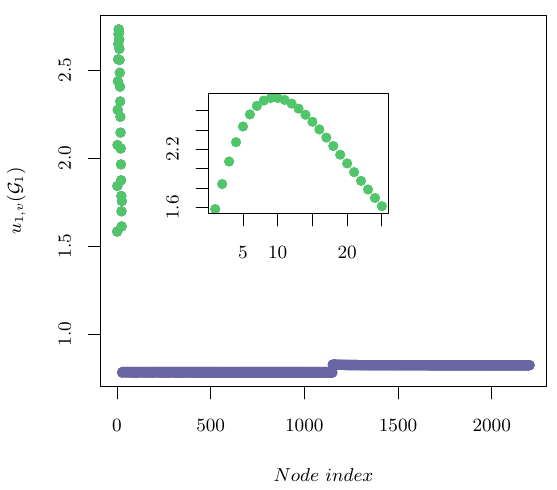}
    \caption{}
    \end{subfigure}
\hfill
    \begin{subfigure}[c]{0.45\textwidth}
    \centering
    \includegraphics[scale=0.8]{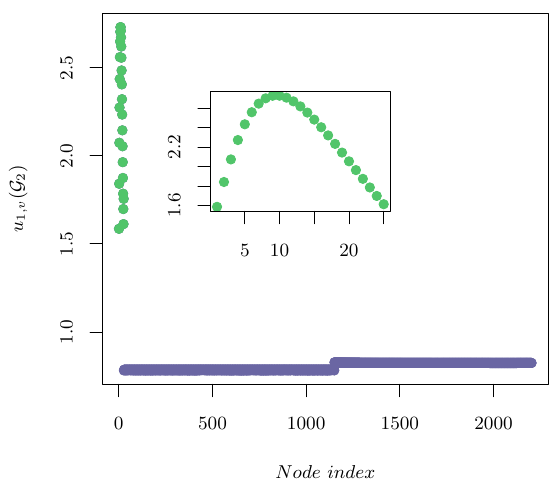}
    \caption{}
    \end{subfigure}
    \caption{Curvatures $u_1(v)$ for the Paley graphs after applying the vertex identification/individualization algorithm, which separates previously indistinguishable vertices and enables recovery of an isomorphic correspondence via sorted curvature values.} 
\label{fig:payblowupcur}
\end{figure}

We show the results of the vertex-individualization construction in Figure~\ref{fig:payblowup}. 
We then compute curvatures associated with the resultant graphs shown in Figure~\ref{fig:payblowupcur}.
They exhibit clear vertex separation, allowing us to identify an isomorphic mapping between vertices using sorted curvature values. 

\subsection{Known benchmark graphs}\label{sec:benchmarks}

We evaluate the algorithm on curated benchmark instances drawn from the repository hosted at
\url{https://pallini.di.uniroma1.it/Graphs.html}.
These graphs include many classical stress tests for refinement- and spectrum-based isomorphism procedures (e.g., highly regular families, strongly regular constructions, cages, and projective or algebraic graphs).
The purpose of this section is to complement the randomly generated experiments of the previous section with standardized, named instances that are widely used in the literature.

We emphasize that these experiments are intended to assess robustness across standard graph families rather than to provide a head-to-head performance comparison with highly optimized practical solvers such as Nauty/Traces; such a comparison remains important future work.

Each family’s files in the repository follow a naming pattern: for example, {\tt ag\_*}, {\tt cfi\_*}, {\tt grid\_*}, {\tt latin\_*}, etc., typically with the family code as prefix.
More broadly, limitations of WL-style refinement and related GI discussion appear in \cite{Weisfeiler1968,Dutta2024}. 
Due to running-time and memory-usage constraints, we consider only benchmark graphs from the repository with a bounded number of original vertices and a bounded number of vertices after subdivision. 
All experiments were executed on a machine with an AMD Ryzen 9 5950X 16-Core CPU with 126GB of RAM. 

\subsubsection{Experimental protocol}

\paragraph{Isomorphic-only evaluation.}
Our benchmark evaluation focuses on isomorphic instances.
For each input graph $G$ we generate $R=3$ relabeled copies
$G^{(r)} = \pi_r(G)$ using deterministic permutations $\pi_r$ derived from a fixed global seed.
Each trial therefore consists of a pair $(G, G^{(r)})$ that is guaranteed to be isomorphic.

\paragraph{Success criterion and one-sided correctness.}
A run is counted as successful only if the algorithm outputs an explicit bijection
$f:V(G)\to V(G^{(r)})$ on the original vertices and the mapping is verified by direct
adjacency preservation on the unmodified input graphs.
Because we only accept mappings that pass this final check, the method is one-sidedly correct:
it does not produce false positive isomorphism claims. 

\paragraph{Staged solver configuration (pairs $\rightarrow$ triplets $\rightarrow$ precision).}
Each isomorphic trial is attempted under an escalation schedule reflecting the solver design.
We first run the refinement and augmentation pipeline using pair probing only (triplet probes disabled).
If the run fails, we re-run the same trial with triplet probing enabled as a fallback.
If the trial still fails, we re-run with more conservative numerical settings to mitigate spectral/curvature
quantization instability by controlling eigenvalue scaling, spectral-dimension stability thresholds, and curvature-coefficient scaling,
respectively. 

\subsubsection{Affine geometry graphs}
\begin{figure}[H] %
    \centering
    \includegraphics[scale=0.9]{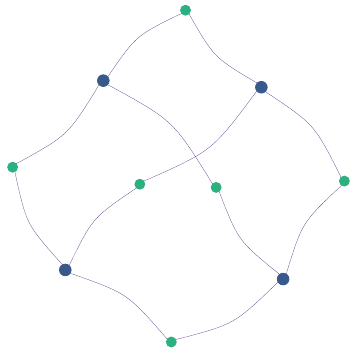}
    \caption{Graph ag2-02 is shown and the vertices are color coded by degree. It has $n = 10$ vertices with degrees 2 and 3. }
    \label{fig:agsample}
\end{figure}

Affine geometry graphs arise from finite affine geometries (e.g.\ points of $\mathrm{AG}(n,q)$) and are highly regular.
They are typically constructed by taking the points of an affine space over $\mathbb{F}_q$ and adding edges according to linear relations or distances in that geometry.
The result is often a strongly regular or distance-regular graph with a large automorphism group (induced by the affine group).
Such regularity makes them challenging for isomorphism tools, as many vertices are combinatorially indistinguishable.
In the repository these graphs have file names beginning with {\tt ag}.
The spectrum typically has few distinct eigenvalues, reflecting its regular structure. Table~\ref{tab:graph_summary_ag} lists all graphs that were successfully solved. A sample graph is shown in Figure \ref{fig:agsample}.

\begin{table*}[ht]
\centering
\begin{tabular*}{\textwidth}{@{\extracolsep{\fill}}|l|c|c|c|c|c|c|}
\hline
\textbf{Graph Name} 
& \multicolumn{2}{c|}{\textbf{Original}} 
& \textbf{\# Subdivisions} 
& \multicolumn{2}{c|}{\textbf{After Subdivision}} 
& \textbf{Triplet Probing} \\
\cline{2-3} \cline{5-6}
& \textbf{$n$} & \textbf{$m$} &  & \textbf{$N$} & \textbf{$M$} & Enabled \\
\hline
ag2-02      & 10  & 12  & 4 & 58  & 60  & No \\
ag2-03     & 21   & 36  & 1  & 57   & 72  & No  \\
ag2-04     & 36  & 80  & 1 & 116  & 160  & No \\
ag2-05     & 55   & 150   & 1  & 205   & 300   & Yes  \\
ag2-07   & 105  & 392  & 1 & 497  & 784  & Yes \\
ag2-08$^\star$   & 136   & 576  & 1 & 712  & 1152  & Yes \\
ag2-09$^\star$   & 171   & 810  & 1 & 981  & 1620  & Yes \\
\hline
\end{tabular*}
\caption{Original graph parameters ($n,m$) and resulting parameters ($N,M$) after edge subdivision. For graphs marked with a star we pick the largest ambiguous class first.}
\label{tab:graph_summary_ag}
\end{table*}

\subsubsection{Cai-F\"urer-Immerman graphs}
\begin{figure}[H] %
    \centering
    \includegraphics[scale=1]{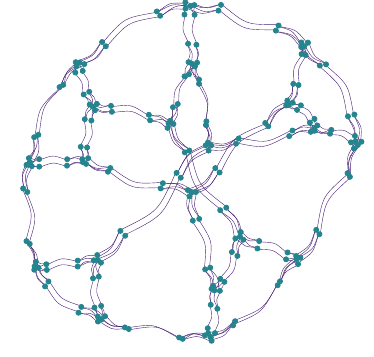}
    \caption{Graph cfi-20 is shown, with $n = 200$ and a homogenous degree of 3.  }
    \label{fig:cfisample}
\end{figure}

This canonical family is built using the CFI construction and is designed to defeat low-dimensional refinement procedures in the Weisfeiler--Leman (WL) hierarchy.
In practice these instances are difficult for purely local refinement rules and require deeper global reasoning; see, e.g., discussions of refinement limits and WL-type methods in \cite{Weisfeiler1968,grohe2015graph,Dutta2024}.
File names in the repository start with {\tt cfi}. Table~\ref{tab:graph_summary_cfi} lists all graphs that were successfully solved, with a sample graph shown in Figure \ref{fig:cfisample}. 

\begin{table*}[ht]
\centering
\begin{tabular*}{\textwidth}{@{\extracolsep{\fill}}|l|c|c|c|c|c|c|}
\hline
\textbf{Graph Name} 
& \multicolumn{2}{c|}{\textbf{Original}} 
& \textbf{\# Subdivisions} 
& \multicolumn{2}{c|}{\textbf{After Subdivision}} 
& \textbf{Triplet Probing} \\
\cline{2-3} \cline{5-6}
& \textbf{$n$} & \textbf{$m$} &  & \textbf{$N$} & \textbf{$M$} & Enabled \\
\hline
cfi-20      & 200  & 300  & 0 & 200  & 300  & No \\
cfi-22     & 220   & 330  & 0  & 220   & 330  & No  \\
cfi-24     & 240  & 360  & 0 & 240  & 360  & No \\
cfi-26     & 260   & 390   & 0  & 260   & 390   & No  \\
cfi-28   & 280  & 420  & 0 & 280  & 420  & No \\
cfi-30   & 300  & 450  & 0 & 300 & 450 & No \\
\hline
\end{tabular*}
\caption{Original graph parameters ($n,m$) and resulting parameters ($N,M$) after edge subdivision.}
\label{tab:graph_summary_cfi}
\end{table*}

\subsubsection{Miyazaki graphs}
\begin{figure}[H] %
    \centering
    \includegraphics[scale=1.2]{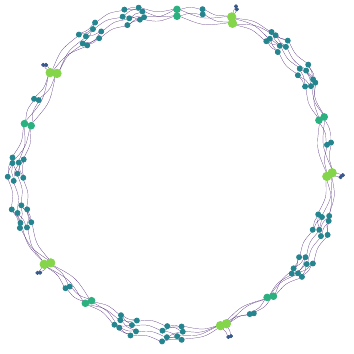}
    \caption{Graph cmz-05 is shown, with $n = 120$ vertices, color coded by degree 2, 3, 4, and 5.  }
    \label{fig:cmzsample}
\end{figure}

Miyazaki graphs are hard, parameterized graphs from Miyazaki-style constructions.
They are used as stress tests because they are very regular and can require considering large-scale swaps to distinguish variants.
The repository uses file names with {\tt cmz} or {\tt mz} (and {\tt mz-aug}, {\tt mz-aug2} for augmented versions).
Table~\ref{tab:graph_summary_mz} lists all graphs that were successfully solved. A sample graph is shown in Figure \ref{fig:cmzsample}.

\begin{table*}[ht]
\centering
\begin{tabular*}{\textwidth}{@{\extracolsep{\fill}}|l|c|c|c|c|c|c|}
\hline
\textbf{Graph Name} 
& \multicolumn{2}{c|}{\textbf{Original}} 
& \textbf{\# Subdivisions} 
& \multicolumn{2}{c|}{\textbf{After Subdivision}} 
& \textbf{Triplet Probing} \\
\cline{2-3} \cline{5-6}
& \textbf{$n$} & \textbf{$m$} &  & \textbf{$N$} & \textbf{$M$} & Enabled \\
\hline
cmz-5      & 120  & 190  & 0 & 120  & 190  & No \\
cmz-6     & 144   & 228  & 0  & 144   & 228  & No  \\
cmz-7     & 168  & 266  & 0 & 168  & 266  & No \\
cmz-8     & 192   & 304   & 0  & 192   & 304   & No  \\
cmz-9   & 216  & 342  & 0 & 216  & 342  & No \\
cmz-10   & 240  & 380  & 0 & 240  & 380  & No \\

mz-2   & 40  & 60  & 1 & 100  & 120  & No \\
mz-4   & 80  & 120  & 0 & 80  & 120  & No \\
mz-6   & 120  & 180  & 0 & 120  & 180  & No \\
mz-8   & 160  & 240  & 0 & 160  & 240  & No \\
mz-10   & 200  & 300  & 0 & 200  & 300  & No \\

mz-aug-2   & 40  & 92  & 1 & 132  & 184  & No \\
mz-aug-4   & 80  & 184  & 0 & 80  & 184  & No \\
mz-aug-6   & 120  & 276  & 0 & 120  & 276  & No \\
mz-aug-8   & 160  & 368  & 0 & 160  & 368  & No \\
mz-aug-10$^\star$   & 200  & 460  & 0 & 200  & 460  & Yes \\

mz-aug2-4   & 96  & 152  & 0 & 96  & 152  & No \\
mz-aug2-6   & 144  & 228  & 0 & 144  & 228  & No \\
mz-aug2-8   & 192  & 304  & 0 & 192  & 304  & No \\
mz-aug2-10   & 240  & 380  & 0 & 240  & 380  & Yes \\
mz-aug2-12$^\star$   & 288  & 456  & 0 & 288  & 456  & Yes \\
\hline
\end{tabular*}
\caption{Original graph parameters ($n,m$) and resulting parameters ($N,M$) after edge subdivision. For graphs marked with a star we pick the largest ambiguous class first.}
\label{tab:graph_summary_mz}
\end{table*}

\subsubsection{Grid graphs}
\begin{figure}[H] %
    \centering
    \includegraphics[scale=1]{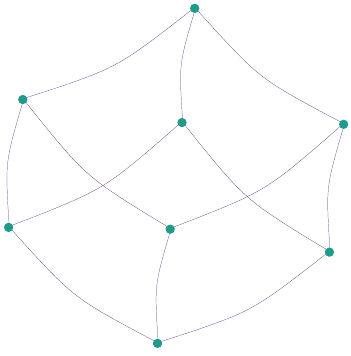}
    \caption{Graph grid-w-3-2 with $n = 8$. }
    \label{fig:gridsample}
\end{figure}

The (Cartesian) grid graphs are formed as the product of two path graphs.
A typical grid graph on $m\times n$ vertices connects each vertex to its neighbors in a square lattice.
They are planar, bipartite, and have many automorphisms (translations and reflections).
In the repository, file names begin with {\tt grid}.
These graphs have simple spectra (being Cartesian products) and are relevant as a baseline. 
Grid graphs with switched edges (grid-sw) are derived by taking a grid graph and swapping a pair of edges in each of several sub-squares.
The operation preserves degree sequences but can obfuscate global structure.
The repository files use prefix {\tt grid-sw}.
Such switched-edge graphs can be harder for refinement algorithms because local neighborhoods remain similar even though global geometry changes.
Table~\ref{tab:graph_summary_grid} lists all graphs that were successfully solved, and a sample graph is shown in Figure \ref{fig:gridsample}.

\begin{table*}[ht]
\centering
\begin{tabular*}{\textwidth}{@{\extracolsep{\fill}}|l|c|c|c|c|c|c|}
\hline
\textbf{Graph Name} 
& \multicolumn{2}{c|}{\textbf{Original}} 
& \textbf{\# Subdivisions} 
& \multicolumn{2}{c|}{\textbf{After Subdivision}} 
& \textbf{Triplet Probing} \\
\cline{2-3} \cline{5-6}
& \textbf{$n$} & \textbf{$m$} &  & \textbf{$N$} & \textbf{$M$} & Enabled \\
\hline
grid-2-5      & 25  & 40  & 1 & 65  & 80  & No \\
grid-2-10     & 100   & 180  & 0  & 100   & 180  & No  \\
grid-2-15     & 225  & 420  & 0 & 225  & 420  & No \\
grid-2-20     & 400   & 760   & 0  & 400   & 760   & No  \\
grid-3-2   & 8  & 12  & 4 & 56  & 60  & No \\
grid-3-3   & 27  & 54  & 1 & 81  & 108  & No \\
grid-3-4   & 64  & 144  & 0 & 64  & 144  & No \\
grid-3-5   & 125  & 300  & 0 & 125  & 300  & No \\
grid-3-6    & 216  & 540  & 0 & 216  & 540  & No \\
grid-3-7   & 343  & 882  & 0 & 343  & 882  & No \\

grid-w-2-5      & 25  & 50  & 1 & 75  & 100  & No \\
grid-w-2-10     & 100   & 200  & 0  & 100   & 200  & No  \\
grid-w-2-15     & 225  & 450  & 0 & 225  & 450  & No \\
grid-w-2-20     & 400  & 800  & 0 & 400  & 800  & No \\
grid-w-3-2     & 8   & 12   & 4  & 56   & 60   & No  \\
grid-w-3-3   & 27  & 81  & 1 & 108  & 162  & No \\
grid-w-3-4   & 64  & 192  & 1 & 256  & 384  & No \\
grid-w-3-5   & 125  & 375  & 0 & 125  & 375  & No \\
grid-w-3-6   & 216  & 648  & 0 & 216  & 648  & No \\
grid-w-3-7  & 343  & 1029  & 0 & 343  & 1029  & No \\
\hline
\end{tabular*}
\caption{Original graph parameters ($n,m$) and resulting parameters ($N,M$) after edge subdivision.}
\label{tab:graph_summary_grid}
\end{table*}

\subsubsection{Hadamard matrix graphs}
Given an $n\times n$ Hadamard matrix $H$, one constructs a Hadamard graph on $4n$ vertices by introducing row-vertices $r_i^\pm$ and column-vertices $c_j^\pm$ and connecting them according to the sign pattern of $H_{ij}$.
This yields a highly regular bipartite graph.
Repository files use prefix {\tt had}. 
One can also apply edge-switching perturbations to Hadamard graphs.
These are labeled with {\tt had-sw}.
Like other switched-edge families, they remain regular, so refinement can stall without additional global cues.
Table~\ref{tab:graph_summary_had} lists all graphs that were successfully solved. 

\begin{table*}[ht]
\centering
\begin{tabular*}{\textwidth}{@{\extracolsep{\fill}}|l|c|c|c|c|c|c|}
\hline
\textbf{Graph Name} 
& \multicolumn{2}{c|}{\textbf{Original}} 
& \textbf{\# Subdivisions} 
& \multicolumn{2}{c|}{\textbf{After Subdivision}} 
& \textbf{Triplet Probing} \\
\cline{2-3} \cline{5-6}
& \textbf{$n$} & \textbf{$m$} &  & \textbf{$N$} & \textbf{$M$} & Enabled \\
\hline
had-1      & 4  & 4  & 12 & 52  & 52  & No \\
had-2     & 8   & 12  & 4  & 56   & 60  & No  \\
had-4     & 16  & 40  & 2 & 96  & 120  & No \\
had-8     & 32   & 144   & 1  & 176   & 288   & No  \\
had-12   & 48  & 312  & 1 & 360  & 624  & No \\
had-16   & 64  & 544  & 1 & 608  & 1088  & No \\
had-20   & 80  & 840  & 1 & 920  & 1680  & Yes \\
had-sw-20 & 80  & 840  & 1 & 920  & 1680  & Yes \\
\hline
\end{tabular*}
\caption{Original graph parameters ($n,m$) and resulting parameters ($N,M$) after edge subdivision.}
\label{tab:graph_summary_had}
\end{table*}

\subsubsection{Latin square graphs}
From a Latin square of order $n$, the Latin square graph has $n^2$ vertices (cells), with adjacency defined by sharing a row, a column, or a symbol.
The result is strongly regular. 
File names start with {\tt latin}.
Variants such as {\tt latin-sw} apply controlled edge modifications intended to preserve coarse invariants while changing global structure.
Table~\ref{tab:graph_summary_latin} lists all graphs that were successfully solved. 

\begin{table*}[ht]
\centering
\begin{tabular*}{\textwidth}{@{\extracolsep{\fill}}|l|c|c|c|c|c|c|}
\hline
\textbf{Graph Name} 
& \multicolumn{2}{c|}{\textbf{Original}} 
& \textbf{\# Subdivisions} 
& \multicolumn{2}{c|}{\textbf{After Subdivision}} 
& \textbf{Triplet Probing} \\
\cline{2-3} \cline{5-6}
& \textbf{$n$} & \textbf{$m$} &  & \textbf{$N$} & \textbf{$M$} & Enabled \\
\hline
latin-2      & 4  & 6  & 8 & 52  & 54  & No \\
latin-3     & 9   & 27  & 2  & 63   & 81  & No  \\
latin-4     & 16  & 72  & 2 & 160  & 216  & No \\
latin-5     & 15   & 150   & 2  & 325   & 450   & No  \\
latin-6   & 36  & 270  & 2 & 576  & 810  & No \\
latin-7   & 49  & 441  & 2 & 931  & 1323  & No \\
\hline
\end{tabular*}
\caption{Original graph parameters ($n,m$) and resulting parameters ($N,M$) after edge subdivision.}
\label{tab:graph_summary_latin}
\end{table*}

\subsubsection{Lattice graphs}
Lattice graphs are general mesh/product graphs.
In the repository ({\tt lattice} files) they refer to regular tiling graphs or Cartesian products in low dimensions.
Names begin with {\tt lattice}.
Table~\ref{tab:graph_summary_lattice} lists all graphs that were successfully solved. 

\begin{table*}[ht]
\centering
\begin{tabular*}{\textwidth}{@{\extracolsep{\fill}}|l|c|c|c|c|c|c|}
\hline
\textbf{Graph Name} 
& \multicolumn{2}{c|}{\textbf{Original}} 
& \textbf{\# Subdivisions} 
& \multicolumn{2}{c|}{\textbf{After Subdivision}} 
& \textbf{Triplet Probing} \\
\cline{2-3} \cline{5-6}
& \textbf{$n$} & \textbf{$m$} &  & \textbf{$N$} & \textbf{$M$} & Enabled \\
\hline
lattice-4      & 16  & 48  & 2 & 112  & 144  & No \\
lattice-5     & 25   & 100  & 2  & 225   & 300  & No  \\
lattice-6     & 36  & 180  & 2 & 396  & 540  & No \\
lattice-7     & 49   & 294   & 2  & 637   & 882   & No  \\
lattice-8   & 64  & 448  & 2 & 960  & 1344  & No \\
\hline
\end{tabular*}
\caption{Original graph parameters ($n,m$) and resulting parameters ($N,M$) after edge subdivision.}
\label{tab:graph_summary_lattice}
\end{table*}

\subsubsection{Additional Paley graphs}
For a prime power $q\equiv 1 \pmod{4}$, the Paley graph of order $q$ has vertex set $\mathbb{F}_q$ and edges $\{a,b\}$ whenever $a-b$ is a quadratic residue in $\mathbb{F}_q$.
These graphs are classical strongly regular examples and are frequently used as structured benchmarks.
Files use prefix {\tt paley}.
Table~\ref{tab:graph_summary_paley} lists all graphs that were successfully solved. 

\begin{table*}[ht]
\centering
\begin{tabular*}{\textwidth}{@{\extracolsep{\fill}}|l|c|c|c|c|c|c|}
\hline
\textbf{Graph Name} 
& \multicolumn{2}{c|}{\textbf{Original}} 
& \textbf{\# Subdivisions} 
& \multicolumn{2}{c|}{\textbf{After Subdivision}} 
& \textbf{Triplet Probing} \\
\cline{2-3} \cline{5-6}
& \textbf{$n$} & \textbf{$m$} &  & \textbf{$N$} & \textbf{$M$} & Enabled \\
\hline
paley-5      & 5  & 5  & 9 & 50  & 50  & No \\
paley-9     & 9   & 18  & 3  & 63   & 72  & No  \\
paley-13     & 13  & 39  & 2 & 91  & 117  & No \\
paley-17     & 17   & 68   & 2  & 153   & 204   & No  \\
paley-25   & 25  & 150  & 2 & 325  & 450  & No \\
paley-29   & 29  & 203  & 2 & 435  & 609  & No \\
paley-37   & 37  & 333  & 2 & 703  & 999  & No \\
\hline
\end{tabular*}
\caption{Original graph parameters ($n,m$) and resulting parameters ($N,M$) after edge subdivision.}
\label{tab:graph_summary_paley}
\end{table*}

\subsubsection{Desarguesian projective plane graphs}
These are point-line incidence (Levi) graphs of finite projective planes of order $q$.
They are bipartite regular graphs of degree $q+1$ with very uniform local structure and large automorphism groups.
File names begin with {\tt pg}.
Such incidence constructions are representative difficult inputs for isomorphism tools; see, e.g., \cite{Gamkrelidze2018} for invariants and discussion in the GI context. 
The {\tt pp} family refers to incidence graphs from small projective planes, including non-Desarguesian examples.
They have the same coarse parameters as the Desarguesian cases but different combinatorial structure.
File names begin with {\tt pp}.
Table~\ref{tab:graph_summary_pp} lists all graphs that were successfully solved. 

\begin{table*}[ht]
\centering
\begin{tabular*}{\textwidth}{@{\extracolsep{\fill}}|l|c|c|c|c|c|c|}
\hline
\textbf{Graph Name} 
& \multicolumn{2}{c|}{\textbf{Original}} 
& \textbf{\# Subdivisions} 
& \multicolumn{2}{c|}{\textbf{After Subdivision}} 
& \textbf{Triplet Probing} \\
\cline{2-3} \cline{5-6}
& \textbf{$n$} & \textbf{$m$} &  & \textbf{$N$} & \textbf{$M$} & Enabled \\
\hline
pg2-2      & 14  & 21  & 2 & 56  & 63  & No \\
pg2-3     & 26   & 52  & 2  & 130   & 156  & No  \\
pg2-4     & 42  & 105  & 2 & 252  & 315  & No \\
pg2-5     & 62  & 186  & 2 & 434  & 558  & Yes \\
pg2-7$^\star$     & 114   & 456  & 2 & 1026  & 1368  & Yes \\

pp-2-1      & 14  & 21  & 2 & 56  & 63  & No \\
pp-3-1     & 26   & 52  & 2  & 130   & 156  & No  \\
pp-4-1     & 42  & 105  & 2 & 252  & 315  & No \\
pp-5-1$^\star$ & 62  & 186  & 2 & 434  & 558  & Yes \\
pp-7-1$^\star$   & 114  & 456  & 2 & 1026  & 1368  & Yes \\
\hline
\end{tabular*}
\caption{Original graph parameters ($n,m$) and resulting parameters ($N,M$) after edge subdivision. For graphs marked with a star we pick the largest ambiguous class first. }
\label{tab:graph_summary_pp}
\end{table*}

\begin{figure}[H] %
    \centering
    \includegraphics[scale=1]{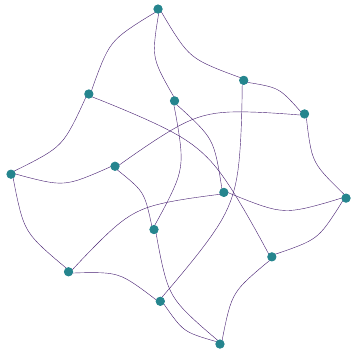}
    \caption{Graph pg2-2 is shown with $n = 14$ and a homogenous degree of 3.  }
    \label{fig:pgsample}
\end{figure}

\subsubsection{Steiner triple system graphs}
\begin{figure}[H] %
    \centering
    \includegraphics[scale=1]{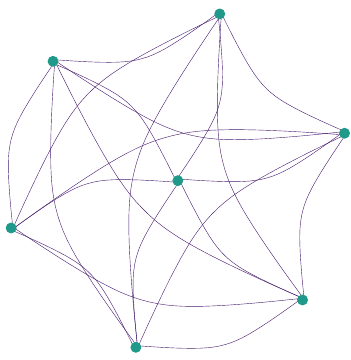}
    \caption{Graph sts\text{-}7 with $n = 7$ is shown with a homogenous degree of 6.}
    \label{fig:stssample}
\end{figure}

From a Steiner triple system, one can form a block-intersection graph: blocks are vertices, and two blocks are adjacent if they share a point.
File names begin with {\tt sts}, with switched variants {\tt sts-sw} also appearing in some benchmark suites.
More broadly, refinement limits and classes where WL-type methods can fail are discussed in \cite{Weisfeiler1968,Dutta2024}.
Table~\ref{tab:graph_summary_sts} lists all graphs that were successfully solved, and a sample representation of the class is shown in Figure \ref{fig:stssample}.

\begin{table*}[ht]
\centering
\begin{tabular*}{\textwidth}{@{\extracolsep{\fill}}|l|c|c|c|c|c|c|}
\hline
\textbf{Graph Name} 
& \multicolumn{2}{c|}{\textbf{Original}} 
& \textbf{\# Subdivisions} 
& \multicolumn{2}{c|}{\textbf{After Subdivision}} 
& \textbf{Triplet Probing} \\
\cline{2-3} \cline{5-6}
& \textbf{$n$} & \textbf{$m$} &  & \textbf{$N$} & \textbf{$M$} & Enabled \\
\hline
sts-7      & 7  & 21  & 3 & 70  & 84  & No \\
sts-9     & 12   & 54  & 2  & 120   & 162  & No  \\
sts-13     & 26  & 195  & 2 & 416  & 585  & No \\
sts-15     & 35   & 315   & 2  & 665   & 945  & No  \\
\hline
\end{tabular*}
\caption{Original graph parameters ($n,m$) and resulting parameters ($N,M$) after edge subdivision.}
\label{tab:graph_summary_sts}
\end{table*}

\subsubsection{Triangular graphs}
The triangular graph $T_n$ is the line graph of the complete graph $K_n$.
Its vertices are the $\binom{n}{2}$ 2-subsets of an $n$-set, with adjacency given by intersection.
In the repository, files are {\tt triang\_n}.
Table~\ref{tab:graph_summary_triang} lists all graphs that were successfully solved. 

\begin{table*}[ht]
\centering
\begin{tabular*}{\textwidth}{@{\extracolsep{\fill}}|l|c|c|c|c|c|c|}
\hline
\textbf{Graph Name} 
& \multicolumn{2}{c|}{\textbf{Original}} 
& \textbf{\# Subdivisions} 
& \multicolumn{2}{c|}{\textbf{After Subdivision}} 
& \textbf{Triplet Probing} \\
\cline{2-3} \cline{5-6}
& \textbf{$n$} & \textbf{$m$} &  & \textbf{$N$} & \textbf{$M$} & Enabled \\
\hline
triang-4      & 6  & 12  & 4 & 54  & 60  & No \\
triang-5     & 10   & 30  & 2  & 70   & 90  & No  \\
triang-6     & 15  & 60  & 2 & 135  & 180  & No \\
triang-7     & 21   & 105   & 2  & 231   & 315   & No  \\
triang-8   & 28  & 168  & 2 & 364  & 504  & No \\
triang-9   & 36  & 252  & 2 & 540  & 756  & No \\
triang-10  & 45  & 360  & 2 & 765  & 1080  & No \\
\hline
\end{tabular*}
\caption{Original graph parameters ($n,m$) and resulting parameters ($N,M$) after edge subdivision.}
\label{tab:graph_summary_triang}
\end{table*}

\subsubsection{Dawar--Yeung graphs}
\begin{figure}[H] %
    \centering
    \includegraphics[scale=1]{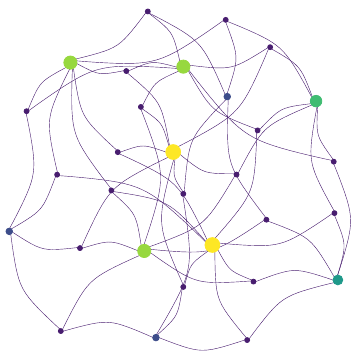}
    \caption{Graph sat-cfi-base-0100-a is shown, with $n = 30$ vertices, color coded by degree $3,4,5,6,7,8$. }
    \label{fig:sat_cfi_dimsample}
\end{figure}

This family ({\tt sat\_cfi}) contains SAT-derived gadget graphs that mimic CFI-style hardness characteristics.
File names are {\tt sat\_cfi\_*}.
Table~\ref{tab:graph_summary_sat} lists all graphs that were successfully solved. A sample graph is shown in Figure \ref{fig:sat_cfi_dimsample}.

\begin{table*}[ht]
\centering
\begin{tabular*}{\textwidth}{@{\extracolsep{\fill}}|l|c|c|c|c|c|c|}
\hline
\textbf{Graph Name} 
& \multicolumn{2}{c|}{\textbf{Original}} 
& \textbf{\# Subd.} 
& \multicolumn{2}{c|}{\textbf{After Subdivision}} 
& \textbf{Triplet Probing} \\
\cline{2-3} \cline{5-6}
& \textbf{$n$} & \textbf{$m$} &  & \textbf{$N$} & \textbf{$M$} & Enabled \\
\hline
sat\_cfi\_base\_0100\_a.dmc     & 30  & 60  & 1 & 90  & 120  & No \\
sat\_cfi\_base\_0100\_b.dmc     & 30  & 60  & 1  & 90   & 120  & No  \\
sat\_cfi\_mult\_0100\_a.dmc     & 100  & 240  & 0 & 100  & 240  & No \\
sat\_cfi\_mult\_0100\_b.dmc     & 100   & 240   & 0  & 100   & 240   & No  \\
sat\_cfi\_base\_0200\_a.dmc   & 60  & 120  & 0 & 60  & 120  & No \\
sat\_cfi\_base\_0200\_b.dmc      & 60  & 120  & 0 & 60  & 120  & No \\
sat\_cfi\_mult\_0200\_a.dmc     & 200   & 480  & 0  & 200   & 480  & No  \\
sat\_cfi\_mult\_0200\_b.dmc     & 200  & 480  & 0 & 200  & 480 & No \\
sat\_cfi\_base\_0300\_a.dmc     & 90   & 180   & 0  & 90   & 180   & No  \\
sat\_cfi\_base\_0300\_b.dmc   & 90  & 180  & 0 & 90  & 180  & No \\
sat\_cfi\_base\_0400\_a.dmc      & 120  & 240  & 0 & 120  & 240  & No \\
sat\_cfi\_base\_0400\_b.dmc     & 120   & 240  & 0  & 120   & 240  & No  \\
sat\_cfi\_base\_0500\_a.dmc     & 150  & 300  & 0 & 150  & 300  & No \\
sat\_cfi\_base\_0500\_b.dmc     & 150   & 300   & 0  & 150   & 300   & No  \\
sat\_cfi\_base\_0600\_a.dmc   & 180  & 360  & 0 & 180  & 360  & No \\
sat\_cfi\_base\_0600\_b.dmc      & 180  & 360  & 0 & 180  & 360  & No \\
sat\_cfi\_base\_0700\_a.dmc     & 210   & 420  & 0  & 210   & 420  & No  \\
sat\_cfi\_base\_0700\_b.dmc     & 210  & 420  & 0 & 210  & 420  & No \\
sat\_cfi\_base\_0800\_a.dmc   & 240  & 480  & 0 & 240  & 480  & No \\
sat\_cfi\_base\_0800\_b.dmc     & 240   & 480   & 0  & 240   & 480   & No  \\
\hline
\end{tabular*}
\caption{Original graph parameters ($n,m$) and resulting parameters ($N,M$) after edge subdivision.}
\label{tab:graph_summary_sat}
\end{table*}

\section{Conclusion and future work}

We have introduced a deterministic, diffusion-driven framework for graph isomorphism that treats vertex identification as the progressive resolution of multi-scale geometric structure.
By extracting curvature-like coefficients from the short-time behavior of the (possibly fractional) graph Laplacian heat kernel and organizing them into BFS-curvature signatures, the algorithm transforms vertex-level diffusion data into structured neighborhood-level invariants.
These invariants induce a refinement hierarchy that propagates geometric information outward from each vertex in a controlled and canonical manner.

When intrinsic diffusion geometry alone does not fully separate vertices, the refinement process is strengthened through bounded, symmetric mechanisms.
These include increasing heat-kernel expansion depth, optional geometric normalization via subdivision to regulate spectral scaling, and structured probing (pair followed by triplet) that exposes higher-order structural distinctions within unresolved equivalence classes. 
If refinement stabilizes without achieving termination, deterministic individualization permanently amplifies residual asymmetry.
All operations are synchronized between the two input graphs and are interpreted exclusively through induced equivalence classes and explicit verification on the original graphs.

The resulting procedure is one-sidedly correct:
any bijection returned by the algorithm is explicitly verified on the original inputs and is therefore guaranteed to be valid. 
The overall refinement hierarchy is deterministic and polynomially bounded under dense linear algebra, with a worst-case upper bound of $O(n^{10})$.
While this bound reflects adversarial cases, empirical behavior is significantly more favorable in the instances tested. 

Across a diverse collection of graph families, including random graphs, highly regular constructions, and standard benchmark instances, the algorithm successfully resolved all tested cases within prescribed refinement limits. 
These experiments suggest that multi-scale diffusion geometry captures structural distinctions that are often inaccessible to raw eigenvalues or purely combinatorial refinement methods.

Overall, the experiments indicate that the diffusion-based refinement pipeline is effective across a broad range of random, structured, and benchmark graph families, including several classes known to be difficult for purely local refinement. At the same time, the present study is intended as an evaluation of correctness and robustness of the framework rather than an optimized performance study; systematic benchmarking against state-of-the-art practical solvers remains an important next step.

Beyond graph isomorphism, this work establishes a concrete algorithmic role for discrete spectral geometry.
Curvature, interpreted as a vector-valued multi-scale diffusion signature rather than a single scalar invariant, functions as a systematic driver of vertex separation. 
Fractional spectral scaling further links discrimination power to intrinsic spectral dimension, providing a principled mechanism for adapting diffusion behavior across graph families.
This geometric viewpoint complements classical combinatorial and group-theoretic techniques and may inform related tasks such as canonical labeling, network alignment, and structural comparison.

Several directions merit further investigation.
From an algorithmic standpoint, optimizing spectral computation, exploring sparse or iterative eigensolvers, and studying scalability to substantially larger graphs are natural next steps.
From a theoretical perspective, characterizing graph classes for which diffusion-based refinement alone suffices, sharpening worst-case bounds, and clarifying the interaction between spectral scaling and automorphism group structure remain open problems.

The present results demonstrate that multi-scale diffusion geometry provides a robust and expressive framework for resolving vertex indistinguishability in graphs. 
Whether this framework can be further strengthened to yield sharper theoretical guarantees remains an open problem, but its algorithmic effectiveness and conceptual coherence suggest a promising direction for continued study.

\bibliographystyle{plain}
\bibliography{arxiv_refs}

@article{DBLP:journals/jsc/McKayP14,
  author       = {Brendan D. McKay and
                  Adolfo Piperno},
  title        = {Practical graph isomorphism, {II}},
  journal      = {J. Symb. Comput.},
  volume       = {60},
  pages        = {94--112},
  year         = {2014},
  url          = {https://doi.org/10.1016/j.jsc.2013.09.003},
  doi          = {10.1016/J.JSC.2013.09.003},
  bibsource    = {dblp computer science bibliography, https://dblp.org}
}

@article{Merris1994,
  author    = {Russell Merris},
  title     = {Laplacian matrices of graphs: a survey},
  journal   = {Linear Algebra and its Applications},
  volume    = {197--198},
  pages     = {143--176},
  year      = {1994},
  doi       = {10.1016/0024-3795(94)90486-3}
}

@article{Fiedler1973,
  author    = {Miroslav Fiedler},
  title     = {Algebraic connectivity of graphs},
  journal   = {Czechoslovak Mathematical Journal},
  volume    = {23},
  number    = {2},
  pages     = {298--305},
  year      = {1973}
}

@article{minakshisundaram1949some,
  title={Some properties of the eigenfunctions of the Laplace-operator on Riemannian manifolds},
  author={Minakshisundaram, Subbaramiah and Pleijel, {\AA}ke},
  journal={Canadian Journal of Mathematics},
  volume={1},
  number={3},
  pages={242--256},
  year={1949},
  publisher={Cambridge University Press}
}

@book{chavel1984eigenvalues,
  title={Eigenvalues in Riemannian geometry},
  author={Chavel, Isaac},
  volume={115},
  year={1984},
  publisher={Academic press}
}

@article{najem2025geometric,
  title={Geometric Features of Higher-Order Networks via the Spectral Triplet},
  author={Najem, Sara and Mrad, Dima and Elsayed, Mohammad},
  journal={arXiv preprint arXiv:2509.04311},
  year={2025}
}

@book{Chung1997,
  author    = {Fan R.~K. Chung},
  title     = {Spectral Graph Theory},
  series    = {CBMS Regional Conference Series in Mathematics},
  volume    = {92},
  publisher = {American Mathematical Society},
  address   = {Providence, RI},
  year      = {1997}
}

@article{Alon1985,
  author    = {Noga Alon and Vitali D. Milman},
  title     = {$\lambda_1$ isoperimetric inequalities for graphs, and superconcentrators},
  journal   = {Journal of Combinatorial Theory, Series B},
  volume    = {38},
  number    = {1},
  pages     = {73--88},
  year      = {1985}
}

@article{Caughman2006,
  author    = {J. S. Caughman and J. J. P. Veerman},
  title     = {Kernels of Directed Graph Laplacians},
  journal   = {Electronic Journal of Combinatorics},
  volume    = {13},
  number    = {R39},
  pages     = {1--13},
  year      = {2006}
}

@article{Raviv2013,
  author    = {Dan Raviv and Ron Kimmel and Alfred M. Bruckstein},
  title     = {Graph Isomorphisms and Automorphisms via Spectral Signatures},
  journal   = {IEEE Transactions on Pattern Analysis and Machine Intelligence},
  volume    = {35},
  number    = {8},
  pages     = {1985--1993},
  year      = {2013},
  doi       = {10.1109/TPAMI.2012.260}
}

@article{DjimaYim2025,
  author    = {Karamatou Yacoubou Djima and Ka Man Yim},
  title     = {Power Spectrum Signatures of Graphs},
  journal   = {arXiv preprint},
  pages     = {arXiv:2503.09660v1},
  year      = {2025},
  url       = {https://arxiv.org/html/2503.09660v1}
}

@book{aho1974design,
  title     = {The Design and Analysis of Computer Algorithms},
  author    = {Aho, Alfred V. and Hopcroft, John E. and Ullman, Jeffrey D.},
  year      = {1974},
  publisher = {Addison-Wesley}
}

@article{luks1982isomorphism,
  title   = {Isomorphism of Graphs of Bounded Valence Can Be Tested in Polynomial Time},
  author  = {Luks, Eugene M.},
  journal = {Journal of Computer and System Sciences},
  volume  = {25},
  number  = {1},
  pages   = {42--65},
  year    = {1982}
}

@article{hopcroft1974linear,
  title   = {A Linear Time Algorithm for Isomorphism of Planar Graphs},
  author  = {Hopcroft, John E. and Wong, James K.},
  journal = {Proceedings of the Sixth Annual ACM Symposium on Theory of Computing},
  pages   = {172--184},
  year    = {1974}
}

@article{grohe2015graph,
  title   = {Graph Isomorphism, Logic, and Polynomial Time},
  author  = {Grohe, Martin},
  journal = {Bulletin of Symbolic Logic},
  volume  = {21},
  number  = {1},
  pages   = {1--45},
  year    = {2015}
}

@article{uehara2005tractable,
  title   = {Tractable and Intractable Instances of the Graph Isomorphism Problem},
  author  = {Uehara, Ryuhei and Toda, Seinosuke and Nagoya, Akira},
  journal = {Theoretical Computer Science},
  volume  = {349},
  number  = {2},
  pages   = {243--256},
  year    = {2005}
}

@article{babai1980canonical,
  title   = {On the Complexity of Canonical Labeling of Strongly Regular Graphs},
  author  = {Babai, L{\'a}szl{\'o}},
  journal = {SIAM Journal on Computing},
  volume  = {9},
  number  = {1},
  pages   = {212--216},
  year    = {1980}
}

@article{babai2016graph,
  title   = {Graph Isomorphism in Quasipolynomial Time},
  author  = {Babai, L{\'a}szl{\'o}},
  journal = {Proceedings of the 48th Annual ACM Symposium on Theory of Computing},
  pages   = {684--697},
  year    = {2016}
}

@article{rozenfeld2007flowers,
  title={Small-world to fractal transition in complex networks: a renormalization group approach},
  author={Rozenfeld, Hern{\'a}n D. and Song, Chaoming and Makse, Hern{\'a}n A.},
  journal={Physical Review Letters},
  volume={98},
  number={12},
  pages={128701},
  year={2007}
}

@article{rozenfeld2010deterministic,
  title={Deterministic scale-free networks},
  author={Rozenfeld, Hern{\'a}n D. and Cohen, Reuven and ben-Avraham, Daniel and Havlin, Shlomo},
  journal={Journal of Physics A: Mathematical and Theoretical},
  volume={43},
  number={42},
  pages={425004},
  year={2010}
}

@article{MasudaPorterLambiotte2020,
  title        = {Random walks and diffusion on networks},
  author       = {Masuda, Naoki and Porter, Mason A. and Lambiotte, Renaud},
  year         = {2020},
  journal      = {arXiv preprint arXiv:1612.03281v3},
  note         = {Survey; reports spectral dimension formulas including $d_s=2\ln(u+v)/\ln(uv)$ for fractal $(u,v)$-flowers.},
}

@article{Hwang2010,
  title   = {Spectral dimensions of hierarchical scale-free networks with shortcuts},
  author  = {Hwang, Sungchul and Yun, Chang-Keun and Kahng, Byungnam and Kim, Doochul},
  journal = {Physical Review E},
  year    = {2010},
  volume  = {82},
  pages   = {056110},
  doi     = {10.1103/PhysRevE.82.056110},
  note    = {Gives spectral dimension results for hierarchical scale-free networks including $(u,v)$-flowers}
}

@article{MasudaRochaEtAl2017,
  title   = {Random walks and diffusion on networks},
  author  = {Masuda, Naoki and Porter, Mason A. and Lambiotte, Renaud},
  journal = {Physics Reports},
  year    = {2017},
  volume  = {716--717},
  pages   = {1--58},
  note    = {Review; summarizes $d_s$ for $(u,v)$-flowers and connects $d_s$ to Laplacian scaling},
  url     = {https://www.cs.cornell.edu/courses/cs6241/2019sp/readings/Masuda-2017-walks.pdf}
}

@article{Bianconi2019,
  title   = {The spectral dimension of simplicial complexes},
  author  = {Bianconi, Ginestra},
  journal = {Physical Review E},
  year    = {2019},
  eprint  = {1910.12566},
  archivePrefix = {arXiv},
  primaryClass  = {cond-mat.stat-mech},
  url     = {https://arxiv.org/pdf/1910.12566},
  note    = {States $\rho(\mu)\sim \mu^{d_S/2-1}$ for small eigenvalues in terms of spectral dimension}
}

@article{Mohar1992,
  title   = {The Laplacian Spectrum of Graphs},
  author  = {Mohar, Bojan},
  journal = {Graph Theory, Combinatorics, and Applications},
  year    = {1992},
  note    = {Survey on Laplacian eigenvalues and their interpretations},
  url     = {https://users.fmf.uni-lj.si/mohar/Papers/Spec.pdf}
}

@article{torres2020simplicial,
  title={Simplicial complexes: higher-order spectral dimension and dynamics},
  author={Torres, Joaqu{\'i}n J and Bianconi, Ginestra},
  journal={Journal of Physics: Complexity},
  volume={1},
  number={1},
  pages={015002},
  year={2020},
  publisher={IOP Publishing},
  doi={10.1088/2632-072X/ab82f5}
}

@article{Sun2009,
  title={A Concise and Provably Informative Multi-Scale Signature Based on Heat Diffusion},
  author={Sun, Jian and Ovsjanikov, Maks and Guibas, Leonidas},
  journal={Computer Graphics Forum},
  volume={28},
  number={5},
  pages={1383--1392},
  year={2009},
  publisher={Wiley},
  doi={10.1111/j.1467-8659.2009.01515.x}
}

@book{Petersen2016,
  title={Riemannian Geometry},
  author={Petersen, Peter},
  edition={3rd},
  publisher={Springer},
  year={2016},
  doi={10.1007/978-3-319-26654-1}
}

@article{Weisfeiler1968,
  title={Reduction of a Graph to a Canonical Form and an Algebra Arising During This Reduction},
  author={Weisfeiler, Boris and Leman, Andrei},
  journal={Nauchno-Technicheskaya Informatsia},
  volume={2},
  pages={12--16},
  year={1968}
}

@article{Reuter2006,
  title={Laplace--Beltrami Spectra as Shape-DNA of Surfaces and Solids},
  author={Reuter, Martin and Wolter, Franz-Erich and Peinecke, Niklas},
  journal={Computer-Aided Design},
  volume={38},
  number={4},
  pages={342--366},
  year={2006},
  doi={10.1016/j.cad.2005.10.011}
}

@article{AlexanderOrbach1982,
  author = {Alexander, S. and Orbach, R.},
  title  = {{Density of states on fractals: “fractons”}},
  journal = {Journal de Physique Lettres},
  volume  = {43},
  pages   = {L625--L631},
  year    = {1982}
}

@article{BurioniCassi1996,
  author = {Burioni, R. and Cassi, D.},
  title  = {{Universal properties of spectral dimension}},
  journal = {Physical Review Letters},
  volume  = {76},
  pages   = {1091--1093},
  year    = {1996}
}

@article{Gamkrelidze2018,
  author    = {Alexander Gamkrelidze and G{\"u}nter Hotz and Levan Varamashvili},
  title     = {New Invariants for the Graph Isomorphism Problem},
  journal   = {CoRR},
  volume    = {abs/1212.3055},
  year      = {2018},
  note      = {\url{https://arxiv.org/abs/1212.3055}}
}

@article{Dutta2024,
  author    = {Sourav Dutta and Arnab Bhattacharya},
  title     = {{\textsf{RSVP}}: Beyond Weisfeiler--Lehman Graph Isomorphism Test},
  journal   = {CoRR},
  volume    = {abs/2409.20157},
  year      = {2024},
  note      = {\url{https://arxiv.org/abs/2409.20157}}
}

@article{Zhang2025FLO,
  author = {Zhang, Mengjie and Lin, Yong and Yang, Yunyan},
  title = {Fractional Laplace operator on finite graphs},
  journal = {Revista Matem\u00e1tica Complutense},
  year = {2025},
  note = {Published online 11 Nov 2025},
  doi = {10.1007/s13163-025-00555-3}
}

@article{Zhang2025FV,
  author = {Zhang, Mengjie and Lin, Yong and Yang, Yunyan},
  title = {Fractional Laplace operator and related Schr\"odinger equations on locally finite graphs},
  journal = {Calculus of Variations and Partial Differential Equations},
  volume = {64},
  pages = {227},
  year = {2025},
  doi = {10.1007/s00526-025-03074-7}
}

@inproceedings{Maskey2023,
  author = {Maskey, Sohir and Paolino, Raffaele and Bacho, Aras and Kutyniok, Gitta},
  title = {A Fractional Graph Laplacian Approach to Oversmoothing},
  booktitle = {Proc. 37th Int. Conf. on Neural Information Processing Systems (NeurIPS)},
  year = {2023}
}

@article{Benzi2020,
  author = {Benzi, Michele and Bertaccini, Daniela and Durastante, Federico and Simunec, Ilaria},
  title = {Non-Local Network Dynamics via Fractional Graph Laplacians},
  journal = {Journal of Complex Networks},
  volume = {8},
  number = {3},
  pages = {cnaa017},
  year = {2020},
  doi = {10.1093/comnet/cnaa017}
}

@article{Weihs2024,
  author = {Weihs, Adrien and Thorpe, Matthew},
  title = {Consistency of Fractional Graph-Laplacian Regularization in Semi-Supervised Learning with Finite Labels},
  journal = {SIAM Journal on Mathematical Analysis},
  volume = {56},
  number = {4},
  year = {2024},
  doi = {10.1137/23M1559087}
}
\end{document}